%% file: paper.tex
\documentclass[a4paper,fleqn,usenatbib]{mnras}

\usepackage{newtxtext}

\usepackage[T1]{fontenc}
\usepackage{ae,aecompl}

\usepackage{natbib,twoopt}
\usepackage{siunitx}
\usepackage[fleqn]{amsmath}
\usepackage{graphicx}
\usepackage{xcolor}
\usepackage{colortbl}
\usepackage{multirow}
\usepackage{scalefnt}

\renewcommand{\vec}[1]{\textbf{#1}}
\usepackage{scalerel}
\usepackage{tikz}
\usetikzlibrary{svg.path}

\definecolor{orcidlogocol}{HTML}{A6CE39}
\tikzset{
  orcidlogo/.pic={
    \fill[orcidlogocol] svg{M256,128c0,70.7-57.3,128-128,128C57.3,256,0,198.7,0,128C0,57.3,57.3,0,128,0C198.7,0,256,57.3,256,128z};
    \fill[white] svg{M86.3,186.2H70.9V79.1h15.4v48.4V186.2z}
                 svg{M108.9,79.1h41.6c39.6,0,57,28.3,57,53.6c0,27.5-21.5,53.6-56.8,53.6h-41.8V79.1z M124.3,172.4h24.5c34.9,0,42.9-26.5,42.9-39.7c0-21.5-13.7-39.7-43.7-39.7h-23.7V172.4z}
                 svg{M88.7,56.8c0,5.5-4.5,10.1-10.1,10.1c-5.6,0-10.1-4.6-10.1-10.1c0-5.6,4.5-10.1,10.1-10.1C84.2,46.7,88.7,51.3,88.7,56.8z};
  }
}

\newcommand\orcidicon[1]{\href{https://orcid.org/#1}{\mbox{\scalerel*{
\begin{tikzpicture}[yscale=-1,transform shape]
\pic{orcidlogo};
\end{tikzpicture}
}{|}}}}

\usepackage{hyperref} 

\title[Stability of Moons Orbiting Kepler-1625 b]{On the Stability of Additional Moons Orbiting Kepler-1625 b}

\author[R. A. Moraes, G. Borderes-Motta, O. C. Winter and J. Monteiro]{%
  R. A. Moraes$^{1}$\thanks{E-mail: ricardo.moraes@unesp.br (RAM)}\orcidicon{0000-0002-4013-8878}\,
  G. Borderes-Motta$^{2}$\thanks{gabriel.borderes@uc3m.es (GBM)}\orcidicon{0000-0002-4680-8414}\, O. C. Winter$^{1}$\thanks{othon.winter@unesp.br (OCW)}\orcidicon{0000-0002-4901-3289}\ and J. Monteiro$^{1}$\thanks{julio.m.santos@unesp.br (JM)}\orcidicon{0000-0002-5152-3560} \\
  $^{1}$ UNESP, Univ. Estadual Paulista - Grupo de Din\^{a}mica Orbital \& Planetologia, Guaratinguet\'{a}, CEP 12.516-410, S\~{a}o Paulo, Brazil\\
  $^{2}$ Bioengineering and Aerospace Engineering Department, Universidad Carlos III de Madrid, Legan\'{e}s, 28911, Madrid, Spain }
\date{Accepted XXX. Received YYY; in original form ZZZ}

\pubyear{2021}

\begin{document}
\label{firstpage}
\pagerange{\pageref{firstpage}--\pageref{lastpage}}
\maketitle


\begin{abstract}
Since it was proposed the exomoon candidate Kepler-1625 b-I changed the way we see satellite systems. Because of its unusual physical characteristics, many questions about the stability and origin of this candidate were raised. Currently, we have enough theoretical studies to assure that if Kepler-1625 b-I is indeed confirmed, it will be stable. The origin of this candidate was also explored. Previous works indicated that the most likely scenario is capture, even though conditions for \textit{in situ} formation were also investigated. In this work, we assume that Kepler-1625 b-I is an exomoon and studied the possibility of an additional, massive exomoon being stable in the same system. To model this scenario we perform N-body simulations of a system including the planet, Kepler-1625 b-I and one extra Earth-like satellite. Based on previous results, the satellites in our system will be exposed to tidal interactions with the planet and gravitation effects due to the rotation of the planet. We found that the satellite system around Kepler-1625 b is capable of harbouring two massive satellites. The extra Earth-like satellite would be stable in different locations between the planet and Kepler-1625 b-I, with a preference for regions inside $25$ $R_p$. Our results suggest that the strong tidal interactions between the planet and the satellites is an important mechanism to assure the stability of satellites in circular orbits closer to the planet, while the 2:1 mean motion resonance between the Earth-like satellite and Kepler-1625 b-I would provide stability for satellites in wider orbits.
\end{abstract}


\input{sec-01} 

\input{sec-02} 

\input{sec-03} 

\input{sec-04} 

\input{conclusion} 

\section*{Acknowledgements}
We thank the anonymous referee for the valuable comments and suggestions that helped improve the quality of this paper. This work was possible thanks to the scholarship granted from the Brazilian Federal Agency for Support and Evaluation of Graduate Education (CAPES), in the scope of the Program CAPES-PrInt, process number 88887.310463/2018-00, Mobility number 88887.583324/2020-00 (RAM). JM thanks the financial support from FAPESP (Grant: 2019/21857-3). RAM, OCW and JM thank the financial supports from FAPESP (Grant: 2016/24561-0) and CNPq (Grant: 305210/2018-1). This research was supported by resources supplied by the Center for Scientific Computing (NCC/GridUNESP) of the S\~{a}o Paulo State University (UNESP).
\section*{Data Availability}
The data underlying this paper will be shared on reasonable request to the corresponding author.
\section*{ORCID iDs}

R. A. Moraes \orcidicon{0000-0002-4013-8878} \href{https://orcid.org/0000-0002-4013-8878}{https://orcid.org/0000-0002-4013-8878}\\
G. Borderes-Motta \orcidicon{0000-0002-4680-8414} \href{https://orcid.org/0000-0002-4680-8414}{https://orcid.org/0000-0002-4680-8414}\\
O. C. Winter \orcidicon{0000-0002-4901-3289} \href{https://orcid.org/0000-0002-4901-3289}{https://orcid.org/0000-0002-4901-3289}\\
J. Monteiro \orcidicon{0000-0002-5152-3560} \href{https://orcid.org/0000-0002-5152-3560}{https://orcid.org/0000-0002-5152-3560}\\



\bibliographystyle{mnras}
\bibliography{ref}

\bsp	
\label{lastpage}
\end{document}

%% file: sec-01.tex
\section{Introduction}
\label{sone}
Natural satellites are common in our Solar System, only two planets, Mercury and Venus, do not host satellites. The variety of satellites around planets is also remarkable. Regarding the size of the satellites, we have bigger satellites, as the Galilean satellite Ganymede which is larger than Mercury and much smaller satellites, such as the inner Uranian satellites. Several special dynamical configurations are also found among the satellites of our Solar System, for example, we have satellites in resonance, most famously the three-body resonant chain, called Laplace resonance, locking the motion of Io, Europa and Ganymede, and co-orbital satellites, as the Janus-Epimetheus case around Saturn. In addition, not only do planets have satellites, in our Solar System satellites are spotted around dwarf-planets, the famous Pluto-Charon binary system for example, and even around asteroids, such is the case of the asteroid 65803 Didymos with its satellite Dimorphos (this system is the target of the NASA's mission DART, launched on 24 November 2021).\par
The abundance and diversity of detected exoplanets suggest the possibility of an equally diverse population of extrasolar satellites, called exomoons. To date no exomoons are confirmed, even with many candidates being proposed \citep{Bennett-etal-2014, Ben-Jaffel-Ballester-2014, Lewis-etal-2015, Hippke-2015, Teachey-etal-2018, Heller-etal-2019, Oza-etal-2019, Fox-Wiegert-2021}. The absence of confirmed exomoons is due mainly to the limitations of the present technology and instruments. Given these constraints, only exomoons with masses higher than $0.1 - 0.5$ Earth masses $(M_{\oplus})$ could be detected by transit \citep{Heller-etal-2014}, currently one of the most promising methods for detection of exomoons. This lower limit for the mass covers bodies that are one order of magnitude more massive than the most massive satellite of our Solar System, Ganymede.\par
Exomoons are a subject of interest in part because they are potentially habitable. \citet{Heller-etal-2014} points out that, because of the amount of Jupiter-like planets inside their respective stellar habitable zone, exomoons could be more likely to harbour biological life than exoplanets. \citet{Zollinger-etal-2017} showed that Mars-like satellites around giant planets in low-mass star systems could be good candidates for habitability under certain assumptions. More recently, \citet{Avila-etal-2021} explored the case of an Earth-like satellite around a "free-floating" Jupiter-like planet and found that a significant amount of liquid water can be retained on the exomoon's atmosphere, thus considering the simulated exomoons as capable of sustaining life. \par
Among the proposed exomoons, the most promising candidate is the one firstly presented by \citet{Teachey-etal-2018}. In this paper, the authors analysed the transit light curves from the \textit{Kepler} telescope of the system Kepler-1625 and found signs of a Neptune-like satellite orbiting a giant planet. However, the analysis presented by \citet{Teachey-etal-2018} was based only on three transits of the planet. To confirm or refute the hypothesis of the exomoon's presence, \citet{Teachey-Kipping-2018} combined the transits from \textit{Kepler} with one more transit signal from the Hubble Space Telescope (HST). The authors found that the new data supported the hypothesis of a moon around Kepler-1625 b. Also, they were able to refine the possible semi-major axis of the candidate and the mass of the planet. \par 
The exomoon candidate hypothesis was contested by \citet{Kreidberg-etal-2019}, in which the authors present a re-analysis of the HST observations using an independent data reduction method. The authors showed that the transit light curve was well fitted by a planet-only model, such as the addition of a moon did not improve the fit. Moreover, \citet{Kreidberg-etal-2019} conclude that the moon signal found by \citet{Teachey-etal-2018} were merely an artifact from data reduction.\par
\citet{Heller-etal-2019} also aimed to collaborate to the debate. The authors presented an alternative interpretation for the \textit{Kepler} and HST data previously analysed. Using an independent reduction method for the HST data and the transits from \textit{Kepler}, \citet{Heller-etal-2019} found a solution that corroborates the planet-moon scenario initially proposed, however, the authors pointed that this result should be carefully regarded because of the systematic errors in the data from \textit{Kepler} and HST. It was suggested that the presence of an exomoon is not a secure detection and that alternative interpretations from the data could be considered, for example, that the later transit by Kepler-1625 b could be due to the presence of a non-detected inclined inner planet. In this case, the planet will be a hot Jupiter with its mass depending on the considered semi-major axis.\par 
More recently, \citet{Teachey-etal-2020} applied different detrending models to the transit light curves presented in \citet{Teachey-Kipping-2018} and found that some reduction techniques are indeed capable to smooth the moon signal. However, from a statistical point of view, the adoption of more flexible detrending models did not provide better evidence than simpler detrending models pointing in favour of a planet-moon scenario. Also, \citet{Teachey-etal-2020} investigated the differences between the analysis of \citet{Teachey-Kipping-2018} and \citet{Kreidberg-etal-2019} and could not find the source of the discrepancies between both results. In this way, the authors argued that the noise properties from both analyses were identical, such as the light curves presented by \citet{Kreidberg-etal-2019} are not better constrained than the ones firstly published.\par
All in all, the current status of the problem is undetermined, with the planet-exomoon model being a valid interpretation for the signal identified in the \textit{Kepler} and HST data.\par
Despite having an exomoon candidate, the Kepler-1625 system seems very ordinary when compared with other exoplanetary systems. It is formed by a G-type star with mass $M_{\star} \sim 1.079$ solar masses $(M_{\odot})$ and radius $R_{\star}\sim 1.793$ solar radius $(R_{\odot})$ \citep{Mathur-etal-2017}. So far, only one planet was detected in the system, Kepler-1625 b. The planet has a semi-major axis of $a_p\sim 0.87$ au \citep{Morton-etal-2016} and radius of $R_p \sim 1.18$ $R_{J}$. The planet's mass is not well constrained, but previous investigations pointed towards a planet with a few Jupiter masses. Recent estimations presented by \citet{Teachey-etal-2020} show a peak probability around $3$ $M_J$. However, as shown in Fig. 10 from \citet{Teachey-etal-2020} the distribution contemplates a broad range of values. The characteristics of the satellite candidate Kepler-1625 b-I are also not well determined. According to \citet{Teachey-Kipping-2018}, the candidate is a Neptune-like body. Several planet-satellite orbital separations were proposed based on different analyses \citep{Teachey-etal-2018, Teachey-Kipping-2018, Martin-etal-2019, Teachey-etal-2020}. In this paper, we will adopt the canonical value of $40$ $R_p$. The exomoon candidate's orbit is usually assumed to be circular. The transit data indicate that the satellite would be in a very inclined configuration. \citet{Teachey-Kipping-2018} found that the inclination of the candidate is $42^{+15}_{-18}$ degrees for a linear detrending, $49^{+21}_{-22}$ degrees for a quadratic detrending and $43^{+15}_{-19}$ degrees for an exponential detrending. As one can see, the inclination varies depending on the model adopted and the error bar for the values is also very significant.\par 
Since it was proposed Kepler-1625 b-I became a favourite target for theoretical models studying its stability \citep{Heller-2018, Quarles-Rosario-Franco-2020, Rosario-Franco-etal-2020} and origin \citep{Heller-2018, Hamers-Zwart-2018, Hansen-2019, Moraes-Vieira-Neto-2020}. From the stability point of view, the satellite candidate is possible and stable since its orbit is well inside the planet's Hill sphere, but not close enough to be dragged towards the planet due to tidal forces. Meanwhile, the origin of this candidate remains mysterious. Most of the papers focus on capture mechanisms for its origin, which is a fair assumption based on the satellite's size, mass and inclination, \citet{Moraes-Vieira-Neto-2020} found that the formation in a circum-planetary disc is also possible, the authors set a lower limit for the number of solids needed for the formation of such massive body and studied its post-formation evolution. Their conclusion points towards \textit{in situ} formation in a supermassive circum-planetary disc. However, the origin of the material was not discussed in their paper.\par 
Based on the features of our Solar System, systems with multiple satellites are expected to be abundant around giant planets. The question is that if this pattern holds for exomoons around giant planets and how big these bodies could be. Many authors studied the stability of single planet-sized satellites \citep{Heller-etal-2014, Heller-Pudritz-2015a, Heller-Pudritz-2015b, Barr-2016}, most of them showing that Mars-like exomoons could be stable around Super Jovian planets. However, is this true for systems with multiple satellites?\par
\citet{Kollmeier-Raymond-2019} and \citet{Rosario-Franco-etal-2020} somewhat explored this issue when they discussed the feasibility of the moon's moons or submoons, around Kepler-1625 b-I. Even though submoons are not found in our Solar System, the size and orbital characteristics of Kepler-1625 b-I motivated \citet{Kollmeier-Raymond-2019} to study the stability around the satellite candidate and in which conditions a submoon would be stable. The authors found that the existence of submoons strongly depends on the mass of the host moon and its orbital separation from the planet, because of the intense tidal interactions between the submoon and the moon. They found that submoons are possible around Kepler-1625 b-I and that the largest possible submoon would have the Vesta-to-Ceres size. Later \citet{Rosario-Franco-etal-2020} found that submoons around Kepler-1625 b-I will orbit inside $0.33$ Hill radius of the satellite (see their Table 1 for more details) and the mass of the stable submoons would be $70\%$ of the mass of the asteroid Vesta.\par
\citet{Moraes-Vieira-Neto-2020} showed that even in an extreme system like Kepler-1625 b, multiple satellites could be stable for long periods in certain locations around the host planet. However, the authors did not explore in detail the location of these bodies or the gravitational interaction between the surviving satellites.\par
\citet{Kipping-2021} identified a phenomenon that would be produced by the presence of one exomoon over the Transit Timing Variation (TTV) of an exoplanet, the "exomoon corridor". According to the author, the analysed TTVs induced by an exomoon will peak near two cycles. In addition, the author estimated that almost half of the exomoons would manifest themselves in short TTVs periods of two to four cycles, regardless of the exomoon's semi-major axis distribution (see Fig. 1 from \citet{Kipping-2021}). If this is the case, the exomoon corridor phenomenon will help to distinguish the TTVs induced by exomoons from the ones induced by planet-planet interactions. However, the exomoon corridor was firstly thought only for one exomoon around a planet. \par
A recent study by \citet{Teachey-2021} expanded the exomoon corridor method to systems with multiple satellites and found that such effect holds for systems with up to five satellites in both resonant and non-resonant configurations.\par
Motivated by the aforementioned works, we propose to investigate in more detail the regions around the planet Kepler-1625 b considering the presence of the satellite Kepler-1625 b-I, searching for orbits where other massive satellites would be stable.\par
We refer to "stable orbits" in our systems as orbits where a secondary satellite can be placed and survive to the gravitational effects of the planet and the primary satellite (Kepler-1625 b-I), and the non-gravitational effects, such as tidal interactions with the central planet and the rotational deformation on the shape of the host body. We chose to take into account tides raised by the planet into the satellites and vice-versa and the rotational flattening effects from the planet, such effect will be modelled by an estimation of the $J_2$ coefficient of the body \citep{Hussmann-etal-2019, Moraes-Vieira-Neto-2020}. \par
The paper is organized as follows. In Section \ref{stwo}, we detail the characteristics of the system, the equations of motion and a deeper discussion about the tidal and the rotational flattening models. Our models and results regarding the stability of the satellites are presented in Section \ref{sthree}, the analysis of the dynamical evolution of our system is shown in Section \ref{sfour}, and in Section \ref{conclusion} we summarize our findings and draw our conclusions.

%% file: sec-02.tex
\section{Model}\label{stwo}
Our main goal is to find stable orbits for massive satellites around Kepler-1625 b, considering that the satellite candidate Kepler-1625 b-I is already present. In this case, we will refer to Kepler-1625 b-I as "primary satellite", while the satellite for which we will study the stability will be identified as "secondary satellite". \par
The study proposed here is supported by previous studies targeting the possibility of finding multiple extrasolar satellites \citep{Heller-etal-2016, Moraes-Vieira-Neto-2020, Teachey-2021}. Also, from the point of view of origins of Kepler-1625 b-I, three scenarios were proposed so far: tidal capture by the planet \citep{Hamers-Zwart-2018}; "pull-down capture" from a coorbital configuration \citep{Hansen-2019}; \textit{in situ} formation \citep{Moraes-Vieira-Neto-2020}. In all cases, the hypothesis of having more satellites in the system is plausible. In the case of capture, Kepler-1625 b-I could have destroyed an ancient family of satellites around the host planet, pushed the ancient exomoons to closer orbits or scattered those satellites in wide eccentric orbits, somewhat similar to the scenario proposed for the capture of Triton on the Neptune system \citep{Agnor-Hamilton-2006}. In the case of \textit{in situ} formation, \citet{Moraes-Vieira-Neto-2020} identified surviving satellites that formed at the same time as Kepler-1625 b-I and are stable in inner orbits.\par 
Because of many uncertainties regarding the values for the mass, radius and semi-major axis of both Kepler-1625 b and Kepler-1625 b-I, for this paper we will consider some canonical values, as summarized in Table \ref{tab:properties}, where the mass and the radius of the satellite are given in terms of mass and radius of Neptune, $M_{Nep}$ and $R_{Nep}$, respectively. Since the information about the eccentricity and inclination of Kepler-1625 b-I are uncertain, we will consider the primary satellite to be in a circular and coplanar orbit around the central planet. Even though the analysis of the transits of Kepler-1625 b presented by \citet{Teachey-Kipping-2018} found that the exomoon candidate could be in a very inclined orbit, thus far it is not possible to know if the planet is tilted on its axis, such that the exomoon's orbit is coplanar with its equator, or if the exomoon candidate is indeed significantly inclined with respect to the planet's equator. In this way, here we opted for the simpler case where the exomoon is coplanar to the planet's equator and leave the inclined case for future explorations.\par
For the secondary satellite, we opted to work with an Earth-like satellite (Tab. \ref{tab:properties}) because of the limitation imposed on the size of detectable satellites. In addition to Earth-like satellites, Ganymede and Mars-like satellites were also tested in the early stages of our study. However, the differences in the size of these three types of satellites were not reflected in our results, therefore we will only show the models with Earth-like bodies as the secondary satellites.
\par
\begin{table*}
\caption[Properties]{Canonical values of mass, radius and semi-major axis adopted for the planet Kepler-1625 b and the satellite Kepler-1625 b-I.}
\begin{tabular}{cccc}\hline 
                & Mass	          & Radius 	       & Semi-major axis        	\\ \hline	
 Kepler-1625 b   & $3$ $M_J$      & $1.18$ $R_J$   & $0.87$ ua                 \\
 Kepler-1625 b-I & $1$ $M_{Nep}$ & $1$ $R_{Nep}$  & $40$ $R_p$                                                        \\ \hline
\end{tabular}
\label{tab:properties}
\end{table*}

\subsection{Equations of Motion}
To describe the motion of the primary and secondary satellites around the planet we will use a coordinate system centred on the planet. The equations of motion of a satellite $j$ with mass $M_j$ at a distance $r_j$ from the central body are given by

\begin{align}\label{eq:motion1}
&\ddot{\vec{r}}_j = -G(M_p+M_j)\dfrac{\vec{r}_j}{|\vec{r}_j|^3} - GM_i\dfrac{\vec{r}_j-\vec{r}_i}{|\vec{r}_j-\vec{r}_i|^3} -  GM_i\dfrac{\vec{r}_i}{|\vec{r}_i|^3}\nonumber \\ &+\dfrac{\vec{F}^T_j + \vec{F}^R_j}{M_j} +\dfrac{\vec{F}^T_j + \vec{F}^R_j}{M_p} + \dfrac{\vec{F}^T_i + \vec{F}^R_i}{M_p}, 
\end{align}
where $j = 1, 2$ and $i= 1, 2$ with $j\neq i$, are the satellites index, $1$ for the primary and $2$ for the secondary. $G$ is the gravitational constant, $r_{j}$ is the planet to $j$-th satellite distance, $M_j$ and$ M_p$ are the masses of the satellite $j$ and the central planet, $\vec{F}^T_j$ and $\vec{F}^R_j$ are, respectively, the tidal and rotational flattening forces acting on the satellite $j$ (these effects are explained on subsection \ref{subsec:tidal}). The terms on the right-hand side (RHS) of Eq. \ref{eq:motion1} are divided into gravitational and tidal/rotational terms. The first three terms are respectively, the gravitational force from the planet on satellite $j$, the mutual gravity interaction between the satellites and an indirect term due to the choice of the coordinate system. The next three terms are the tidal and rotational flattening accelerations due to the bulge induced by the planet on the satellite $j$, the bulge induced by the satellite $j$ on the planet and the bulge induced by the satellite $i$ on the planet, respectively. One should notice that the satellites do not raise tides in each other and the tidal bulge induced in each satellite do not affect the motion of the other, thus the fourth term on the RHS of Eq. \ref{eq:motion1} does not have a similar term for the satellite $i$ (Fig. \ref{fig:bulge}). \par
As one can see from Eq. \ref{eq:motion1}, the presence of the star is neglected. As shown by \citet{Moraes-Vieira-Neto-2020}, for the distances we are considering for the satellites, the effects of the star can be safely ignored, such as the planet will be the gravitationally dominant body.

\subsection{Tidal and Rotational Flattening Models}\label{subsec:tidal}

In addition to the gravitational interaction between the bodies in the system, we also assume tidal forces and rotational flattening as sources of perturbations on the system.\par

\subsubsection{Tidal Model}
\begin{figure}
\begin{center}
\includegraphics[height=0.57\columnwidth, width=\columnwidth]{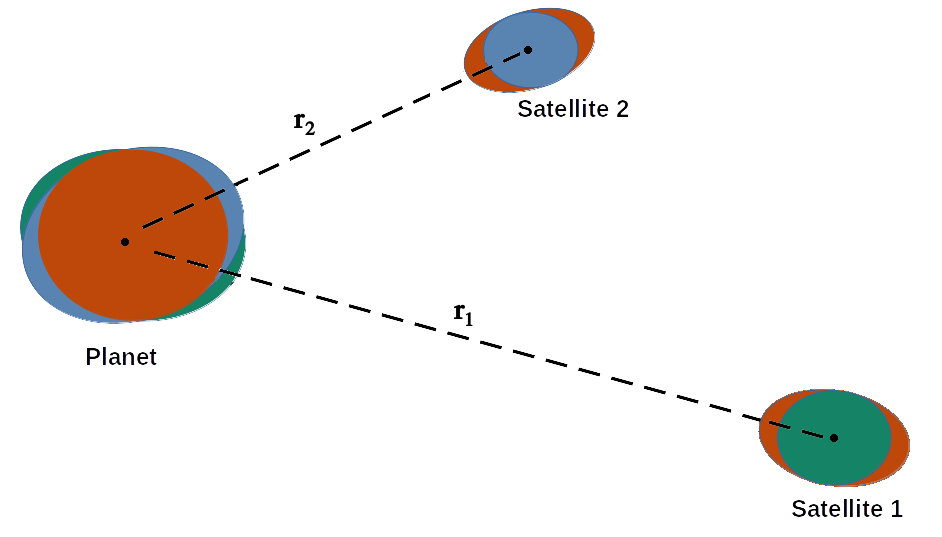}
\caption[Tidal Bulge]{Illustration of how tides are considered in this paper. The central planet exerts tides on each satellite independently and each satellite also exerts tides on the planet, while satellites do not generate tides in each other.}
\label{fig:bulge}
\end{center}
\end{figure}

To proper model the tides in the system, we consider the constant time lag formalism proposed in \cite{Mignard-1979} and \citet{Hut-1981} (see also \citet{Bolmont-etal-2015} and the references therein). In our models we consider that the central planet will generate a tidal bulge in each satellite and each orbiting satellite will also generate a tidal bulge in the central planet, while the satellites do not generate tides in each other (Fig. \ref{fig:bulge}). Also, the tidal bulges induced by the planet on one satellite does not affect the motion of the other. On the other hand, all the bulges generated by the satellites on the planet affect the motion of all the orbiting bodies.\par
To describe the tidal components on the acceleration we will adopt the notation used in \citet{Bolmont-etal-2015}. Thus, we present the tidal forces acting on the satellite $j$ decomposed in radial $(\vec{F}^T_{rad})$ and orthogonal $(\vec{F}^T_{ort})$ terms as follows, 
\begin{align}\label{eq:fi}
 &\vec{F}^T_j = \vec{F}^T_{rad} + \vec{F}^T_{ort} \nonumber \\
 &= \left[F^T_{rad,j} + F^T_{rad,p}+\left(F^T_{ort,j}+F^T_{ort,p}\right)\dfrac{\vec{v}_j\cdot\mathbf{\hat{r}}_j}{|\vec{r}_j|}\right]\mathbf{\hat{r}}_j \nonumber \\
 &+ \left[F^T_{ort,j}\left(\mathbf{\Omega}_j - \dot{\mathbf{\theta}}_j\right) + F^T_{ort,p}\left(\mathbf{\Omega}_p - \dot{\mathbf{\theta}}_j\right)\right]\times \mathbf{\hat{r}}_j,
\end{align}
where $\vec{v}_j$ and $\mathbf{\Omega}_j$ are the velocity and the rotation vectors of the satellite $j$, $\mathbf{\Omega}_p$ is the rotation vector of the planet, and $\mathbf{\hat{r}}_j = \vec{r}_j/|\vec{r}_j|$ is a unit vector in the radial direction and $\mathbf{\dot{\theta}}_j$ is the instantaneous orbital angular velocity vector, which is a colinear vector with the orbital angular momentum of the satellite $j$ ($\theta_j$ is the true anomaly of the respective satellite). Here, we consider the time evolution of the rotation vectors of the planet and the satellites, $\mathbf{\Omega}_p$ and $\mathbf{\Omega}_j$ respectively, through the integration of the torque equations given by the conservation of the total angular momentum (see the sections 2.4.2 and 2.5 of \citet{Bolmont-etal-2015} for a complete analytic development of such equations). In our model, we are neglecting the evolution of the bodies, such as the physical radius and the radius of gyration of the planet and the satellites are kept constant, thus simplifying the evolution of the rotation vectors.\par
The terms $F^T_{rad,j}$ and $F^T_{rad,p}$ refer to the radial tides while $F^T_{ort,j}$ and $F^T_{ort,p}$ are related to the orthogonal tides. The subscripts $j$ and $p$ are related to tides from the satellites $j$ and the planet $p$, respectively. These terms were based on Eq. 5 from \citet{Bolmont-etal-2015} and given by,
\begin{align}\label{eq:terms}
&F^T_{rad,j} = -3\dfrac{GM_p^2 k_{2,j}R_j^5}{|\vec{r}_j|^7}\left(1 + 3\dfrac{|\vec{v}_j|}{|\vec{r}_j|}\tau_j\right),\\ \nonumber
&F^T_{rad,p} = -3\dfrac{GM_j^2 k_{2,p}R_p^5}{|\vec{r}_j|^7}\left(1 + 3\dfrac{|\vec{v}_j|}{|\vec{r}_j|}\tau_p\right),\\ \nonumber
&F^T_{ort,j} = 3\dfrac{GM_p^2R_j^5}{|\vec{r}_j|^7}k_{2,j}\tau_j,\\ \nonumber
&F^T_{ort,p} = 3\dfrac{GM_j^2R_p^5}{|\vec{r}_j|^7}k_{2,p}\tau_p,
\end{align}
where $R_j$ and $R_p$ are the radii of the satellite $j$ and the planet, respectively, $k_{2,j}$ and $\tau_j$ are the second-order potential Love number and the constant time lag of the satellite $j$, while $k_{2,p}$ and $\tau_p$ have the same meaning, but for the central planet.\par
The tides have the effect to damp the eccentricity of the orbiting body while affecting its migration. The direction of the migration depends primarily on the eccentricity of the orbiting body and its semi-major axis.\par
When the satellite has zero eccentricity and zero obliquity, its orbit is perfectly synchronized with the central planet, i.e. the satellite's angular velocity is equal to the planet's rotation velocity, thus only the tides of the planet are present. For small non-zero eccentricities, the tides from the central planet are dominant over the tides of the satellites, in this case, the direction of the migration of the satellites depends on their orbital position: if a satellite is located beyond its corotation radius $(r_{corot})$ then its angular velocity is smaller than the planet's rotation rate and the satellite migrates outwards, and if the satellite is located inside $r_{corot}$ its angular velocity is greater than the planet's rotation rate, thus the satellite migrates inwards. For a high eccentric satellite, the tides from the satellite dominate and the satellite will always migrate inwardly \citep{Leconte-etal-2010, Sanchez-etal-2020}. The rate at which the semi-major axis of a satellite in an eccentric satellite varies is given by \citep{Leconte-etal-2010, Hansen-2010, Sanchez-etal-2020}
\begin{align}\label{eq:dadt}
\dfrac{1}{a}\dfrac{da}{dt} = &- \dfrac{1}{t_{sat}}\left[f_1(e) - \dfrac{\Omega_{sat}}{n}f_2(e)\right]\nonumber \\
&- \dfrac{1}{t_{p}}\left[f_1(e) - \dfrac{\Omega_{p}}{n}f_2(e)\right],
\end{align}
with $n$ being the mean motion of the satellite, $t_{sat}$ and $t_p$ are the dissipation timescales for circular orbits of the satellite and the planet, respectively, given by,
\begin{align}\label{eq:timescales}
&t_{sat} = \dfrac{1}{6}\dfrac{1}{G k_{2,sat}\tau_{sat}}\dfrac{M_{sat}}{M_p(M_{sat}+M_p)}\dfrac{a^8}{R_{sat}^5}, \nonumber \\
&t_{p} = \dfrac{1}{6}\dfrac{1}{G k_{2,p}\tau_{p}}\dfrac{M_{p}}{M_{sat}(M_p+M_{sat})}\dfrac{a^8}{R_{p}^5},
\end{align}
 $f_1(e)$ and $f_2(e)$ are functions of the eccentricity of the satellite that appear for eccentric orbits, given by Eq. \ref{eq:na1na2}.
\begin{align}\label{eq:na1na2}
&f_1(e) = \dfrac{1+(31/2)e^2+(255/8)e^4+(185/16)e^6+(25/64)e^8}{(1-e^2)^{15/2}}, \nonumber \\
&f_2(e) = \dfrac{1+(15/2)e^2+(45/8)e^4+(5/16)e^6}{(1-e^2)^{6}}.
\end{align}
For a more extensive analysis on the secular evolution of the semi-major axis, eccentricity and spin of a body being tidally disturbed see \citet{Leconte-etal-2010}, \citet{Hansen-2010} \citet{Bolmont-etal-2011}, \citet{Bolmont-etal-2013} and \citet{Sanchez-etal-2020}.  \par
As mentioned before in all our models we will consider the central planet, the primary satellite Kepler-1625 b-I and a secondary Earth-like satellite. To proper account for tides in the system, some important parameters must be set, namely the rotation period, the second-order potential Love number and the constant time lag (hereafter mentioned as tidal parameters) for each body. For simplicity, we will consider for the central planet, Kepler-1625 b, to have tidal parameters equal to the ones of Jupiter (parameters taken from \citet{Bolmont-etal-2015}), this choice is justified because the planet has a size similar to Jupiter's and $3$ $M_J$. For a more massive planet $(\ge 10\;M_J)$, Brown-dwarf parameters could be considered. For Kepler-1625 b-I we follow \citet{Tokadjian-Piro-2020} and set the potential Love number and constant time lag to describe an ice-giant planet like Neptune, $k_{2,1} = 0.34$ and $\tau_1 = 0.766$ $s$. Also, the rotation period of the satellite was taken as $16$ hours, the same as the planet Neptune. As the primary satellite is located farther out in the system, where the tides from the planet are weaker, a slower or faster rotating body will produce similar results. In the case of the secondary satellite, we are imposing it to be an Earth-like body with the same tidal parameters as the Earth (values taken from \citet{Bolmont-etal-2015}), we also tested an Earth-like satellite rotating faster ($18$ hours) and slower ($30$ hours), however, the results were very similar to the ones obtained with the typical rotation period of $24$ hours. A summary with the values of the tidal parameters for each body is presented in Tab. \ref{tab:tidal-parameters}.
\begin{table*}
\caption[Tidal Parameters]{Tidal parameters for the planet, Kepler-1625 b, the satellite Kepler-1625 b-I and the secondary satellite. Each column presents, the body, the rotation period, the second-order potential Love number and the constant time lag \citep{Bolmont-etal-2015, Tokadjian-Piro-2020}.}
\begin{tabular}{cccc}\hline 
                        & Period	          & $k_{2}$ 	       & $\tau$        	\\
                        & hour	              &      	           & $s$        	\\ \hline  
 Kepler-1625 b           & $10$              & $0.380$            & $1.842 \times10^{-3}$   \\
 Kepler-1625 b-I         & $16$              & $0.340$            & $0.766$                \\ 
 Secondary satellite    & $24$              & $0.305$            & $698$                         \\ \hline
\end{tabular}
\label{tab:tidal-parameters}
\end{table*}
\subsubsection{Rotational Flattening Model}
Following \citet{Hussmann-etal-2019} and \citet{Moraes-Vieira-Neto-2020} we chose to include in our models the effects of the nonsphericity of the simulated bodies due to their rotation. \par 
Besides the distortions on the surface of a body due to tides, the shape of this body will also change due to its rotation. The gravitational effects of this rotational deformation will change its gravitational field, thus affecting all the bodies in the system, especially the ones in closer orbits \citep{Murray-1999}.\par
Here, we will consider the torques exerted by the planet over the satellites and the torques exerted by the satellites over the planet, while torques from one satellite are not felt by the other satellite. These gravitational effects will precess the orbit of eccentric satellites, thus causing the mean eccentricity and mean obliquity of the satellites to change \citep{Bolmont-etal-2015}.\par
We will model the rotation deformation of a body using the parameter $J_2$, which defines how oblate instead of spherical a body is. For a given satellite $j$, $J_{2,j}$ can be calculated as \citep{Bolmont-etal-2015} \begin{align}\label{eq:J2}
    J_{2,j} = k_{2f,j}\dfrac{\Omega_j^2R_j^3}{3GM_j}
\end{align}
where $k_{2f,j}$ is the second-order fluid Love number of the satellite. The fluid Love number can be defined as the potential Love number for a perfect fluid body. For simplicity, we will consider the fluid Love number to be equal to the potential Love number \citep{Bolmont-etal-2015}, for a complete discussion about the differences between these two parameters see \citet{Correia-Rodriguez-2013} and the references therein. Eq. \ref{eq:J2} can be used to calculate $J_{2,p}$ for the central planet simply exchanging the subscript "$j$" for "$p$".\par
As in the case of the tidal model, we will adopt a notation similar to \citet{Bolmont-etal-2015} and describe the rotational flattening force of a central planet and a satellite $j$ over a satellite $j$ as,
\begin{align}\label{eq:rotation}
&\vec{F}_j^R = -\dfrac{3GM_jM_p}{2|\vec{r}_j|^5}\left(J_{2,j}R_j^2+J_{2,p}R_p^2\right)\vec{r}_j\nonumber \\
&+\dfrac{15GM_jM_p}{2|\vec{r}_j|^7}\left(J_{2,j}R_j^2\dfrac{(\vec{r}_j\cdot\mathbf{\Omega}_j)^2}{|\mathbf{\Omega}_j|^2} + J_{2,p}R_p^2\dfrac{(\vec{r}_j\cdot\mathbf{\Omega}_p)^2}{|\mathbf{\Omega}_p|^2}\right)\vec{r}_j \nonumber \\
&-\dfrac{3GM_jM_p}{|\vec{r}_j|^5}\left(J_{2,j}R_j^2\dfrac{\vec{r}_j\cdot\mathbf{\Omega}_p}{|\mathbf{\Omega}_p|^2}\right)\mathbf{\Omega}_p \nonumber \\
&-\dfrac{3GM_jM_p}{|\vec{r}_j|^5}\left(J_{2,p}R_p^2\dfrac{\vec{r}_j\cdot\mathbf{\Omega}_j}{|\mathbf{\Omega}_j|^2}\right)\mathbf{\Omega}_j. 
\end{align}
Now that we defined all the forces acting in our system and can compute the total acceleration felt by a satellite $j$ (Eq. \ref{eq:motion1}), we will move to describe our computational methods and our set-ups.

%% file: sec-03.tex
\section{Numerical Simulations}\label{sthree}

Our study will be mainly numerical, thus we start this section presenting the numerical tools used in this work, followed by a discussion about the systems studied and the results of our simulations regarding the stability of the systems. The analysis of the dynamical evolution of the systems is presented in Section \ref{sfour}.

\subsection{Numerical Tools}

In this work we will perform N-body simulations using the numerical package POSIDONIUS\footnote[1]{https://www.blancocuaresma.com/s/posidonius} \citep{Blanco-Cuaresma-Bolmont-2017a}. POSIDONIUS is a new N-body code that allows the user to include dissipative effects such as tides, general relativity, rotational flattening and the evolution of the bodies. The package is the second generation of the widely used MERCURY-T \citep{Bolmont-etal-2015} written in Rust (see \citet{Blanco-Cuaresma-Bolmont-2017b} for the advantages of using Rust in astrophysics) with a package of Python scripts for the implementation of initial conditions and post-processing analysis. This new numerical tool ensures memory safety, can reproduce numerical N-body experiments and improves the integration of the spin when compared with MERCURY-T. POSIDONIUS has been used in recent studies involving the evolution of planets and satellites in scenarios where tides are considered \citep{Bolmont-etal-2019, Bolmont-etal-2020}.\par
POSIDONIUS allows the user to choose between different integration schemes for the time evolution of the equations of motion. Here, we opted to use the IAS15 integration scheme \citep{Rein-Spiegel-2015}. The main reason for this choice is that in our problem we expect to find several close encounters between the satellites and the satellites with the central planet. The IAS15 integrator can solve such close encounters with high accuracy while still being faster than most other high-order integration schemes \citep{Rein-Spiegel-2015}.\par

\subsection{Initial Conditions}

As mentioned before, the systems we are studying are composed of the central planet (Kepler-1625 b), the primary satellite (Kepler-1625 b-I) and a secondary Earth-like satellite. The physical characteristics of the planet and the primary satellite are given in Tab. \ref{tab:properties}, while the secondary satellite will have Earth's radius and mass. The primary satellite will start at $40$ $R_p$ from the planet for all cases and in circular and coplanar orbit. In order to find stable orbits for the secondary satellite we will simulate several cases with the initial semi-major axis of the satellites varying from $a_0 = 5$ to $a_0 = 35$ $R_p$, considering $\Delta a_0 = 2$ $R_p$. For each position we will simulate ten cases, varying the eccentricity of the secondary satellite from $e_0 = 0.0$ to $e_0 = 0.9$, with $\Delta e_0 = 0.1$. The satellites are taken to be coplanar in all cases. The angular orbital elements of both satellites and their obliquities were set to zero. In the case of the obliquity of the satellites, as we are considering the effects of tides and rotation flattening in addition to the usual gravitational interactions, \citet{Bolmont-etal-2015} showed for several cases that the obliquity of orbiting bodies under these dissipative effects decreases to zero very quickly, around $100$ years. We used their results to justify our choice.\par
As we are considering additional effects to our model, we present the tidal parameters of all the bodies in the system in Tab. \ref{tab:tidal-parameters}, recalling that we only consider tides generated by the planet on the satellites and the tides raised by the satellites on the planet, neglecting satellite-satellite tidal interactions. For the gravitational effects due to rotational flattening the "only" parameter needed is $J_2$, which is calculated for each satellite and the planet with Eq. \ref{eq:J2}. Each simulation will be carried for $1$ Myr or until a collision or an ejection is detected. Satellites are ejected from the system if they reach $1$ Hill radius from the planet at any point. If a system survives for $1$ Myr we consider the system as stable, otherwise the system will be classified as unstable.\par
Hereafter we will refer to all systems by the initial semi-major axis and eccentricity of the secondary satellite.

\subsection{Stability}
\begin{figure}
\begin{center}
\includegraphics[height=0.7\columnwidth, width=\columnwidth]{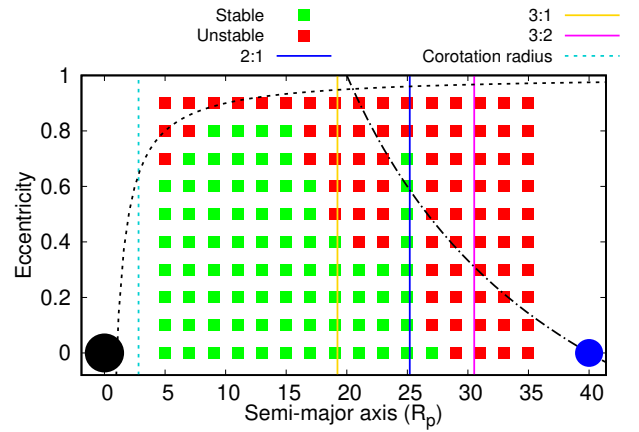}
\caption[Results Simulations]{Grid with the initial conditions. In green, we have the conditions that end up being stable after $1$ Myr and in red the unstable cases. The vertical lines show the approximate location of some mean-motion resonances (MMR): 2:1 (blue line), 3:2 (magenta line) and 3:1 (yellow line). The innermost dashed line shows the initial location of the corotation radius for the Earth-like moons. The black and blue circles represent the central planet and Kepler-1625 b-I, respectively. The black dashed line shows where the satellites' pericentre crosses the radius of the planet and the black dash-dotted line shows where the satellites' apocentre crosses the orbit of the primary satellite.}
\label{fig:a-e-results}
\end{center}
\end{figure}
In the following, we discuss our results regarding the stability of the systems when a secondary satellite is added. Also, we will analyse the fate of the unstable systems.\par 
In Fig. \ref{fig:a-e-results} we present our grid (semi-major axis vs eccentricity) of initial conditions. In green, we show the stable initial conditions and in red the unstable ones. Later we will discuss the orbital evolution and the final configurations of the satellite systems, but for now, is important noting that even very eccentric initial conditions can provide long-term stable satellites due to the effects of tides. As pointed out in the previous section, tides are a strong mechanism to damp the eccentricity of the orbiting bodies.\par
From Fig. \ref{fig:a-e-results} we can see that the systems with the secondary satellite initially placed inside $20$ $R_p$ are more likely to produce stable systems, this is the case because at these locations the gravitational effects of the primary satellite are weaker and the evolution of the systems is dominated by the tidal interaction with the planet. Outside $20$ $R_p$ the effects of the tides are weaker and the gravitational perturbations from the primary satellite are more pronounced. At these locations, we see a decrease in the number of stable systems, especially among the ones with orbits initially eccentric. We point out, that secondary satellites in eccentric orbit are more likely to suffer a close encounter with the primary satellite, resulting in collisions or ejection from the system. In this way, the action of the tides as a mechanism to damp the eccentricity of the secondary satellites is very important to reduce the probability of these close encounters.\par
For systems with the secondary satellite placed at distances greater than $29$ $R_p$ we could not find stability. At these distances the effects of the primary satellite are dominant and close encounters that lead to the loss of the secondary satellites are common (the dash-dotted line in Fig. \ref{fig:a-e-results} shows the conditions for which the satellites' apocentre crosses the orbit of the primary satellite and close encounters are more likely). In addition, we will show later that the overlapping of resonances could create a chaotic region close to the primary satellite.  \par
Also in Fig. \ref{fig:a-e-results} we present the initial location of three mean-motion resonances (MMR), the 3:1 near $19$ $R_p$, the 2:1 around $25$ $R_p$ and the 3:2 close to $31$ $R_p$. These three MMRs play different roles in the stability of the systems. As one can see, the 3:1 MMR is positioned inside a stable region, thus the number of stable systems does not increase/decrease substantially because of the presence of the resonance. On the other hand, the 2:1 MMR is located in a transition region, where the number of stable systems decay. However, the system with secondary satellites starting near this region present a spike in the number of stable systems, which is directly correlated with the presence of the resonance. The outermost resonance we present is the 3:2 MMR slightly inside $31$ $R_p$. This region is too close to the primary satellite to be stable and the MMR resonance does not contribute to the stability of the systems, most of the systems with the secondary satellite starting at $31$ $R_p$ became unstable almost instantaneously. A more detailed study about the role the MMRs play in the systems will be presented in the next section.\par
\begin{figure}
\begin{center}
\includegraphics[height=0.7\columnwidth, width=\columnwidth]{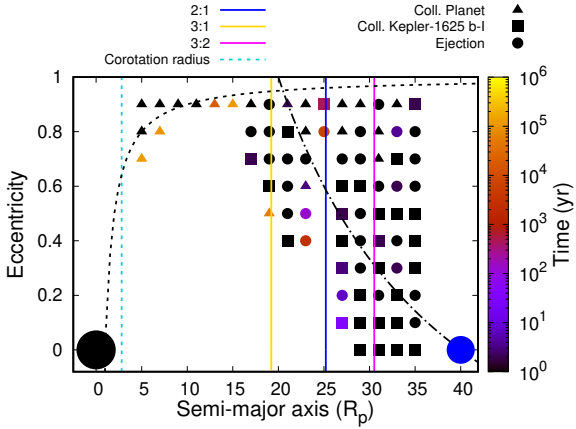}
\caption[Unstable cases]{Grid with the unstable initial conditions of our models. The shape of the points indicates the fate of the satellite: triangles represent collision with the central planet, squares represent collision with the primary satellite (Kepler-1625 b-I) and circles represent ejection from the satellite system. The colour of the points indicates the time when the system became unstable according to the colour bar. The black dashed line shows where the satellites' pericentre crosses the radius of the planet and the black dash-dotted line shows where the satellites' apocentre crosses the orbit of the primary satellite.}
\label{fig:a-e-unstable}
\end{center}
\end{figure}
To investigate the fate of the unstable systems, in Fig. \ref{fig:a-e-unstable} we show a diagram with only the unstable initial conditions, where the triangles represent a collision with the central planet, squares represent a collision with Kepler-1625 b-I and circles represent ejections from the system. The colour bar on the right side indicates the time it took for the system to become unstable and the points marked in black are the conditions that became unstable instantaneously after the simulation started.\par
From Fig. \ref{fig:a-e-unstable} one can see that, as expected, the satellites initially in inner orbits are more likely to collide with the central planet, in the same way, satellites initially in wider orbits are more likely to collide with the primary satellite or be ejected from the system. However, satellites in wider eccentric orbits also collided with the planet, this happens because close encounters with the primary satellite ended with the secondary satellite being pushed inwards, towards the planet and a collision eventually takes place. We found several cases of ejection of the secondary satellite, the nature of the ejection is mainly due to close encounters with the primary satellite (especially for the secondary satellites located at outer regions) or due to a resonant break.\par
\begin{figure*}
\begin{center}
\includegraphics[height=0.47\linewidth, width=\linewidth]{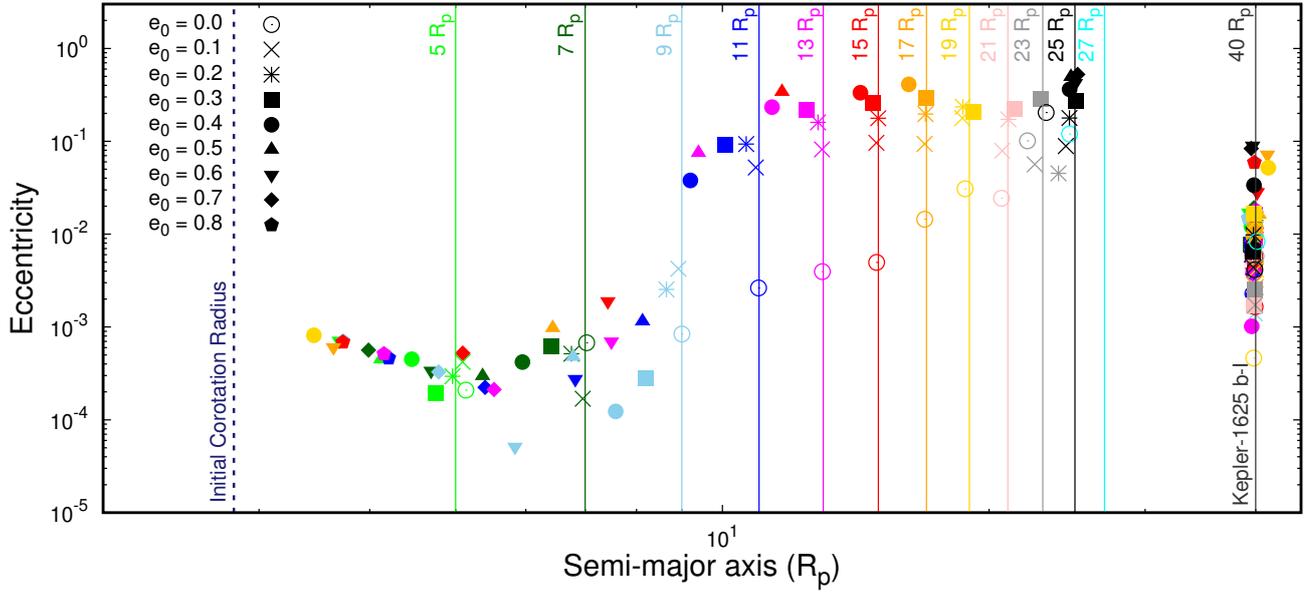}
\caption[Stable cases]{Final semi-major axis vs eccentricity distribution of the stable cases of our models. The models are separated by colour and shape. The colours identify the initial semi-major axis of the satellite $(a_0)$ from $5$ $R_p$ to $35$ $R_p$, with $\Delta a_0 = 2$ $R_p$, as light-green ($5$ $R_p$), green ($7$ $R_p$), light-blue ($9$ $R_p$), blue ($11$ $R_p$), magenta ($13$ $R_p$), red ($15$ $R_p$), orange ($17$ $R_p$), yellow ($19$ $R_p$), pink ($21$ $R_p$), gray ($23$ $R_p$), black ($25$ $R_p$) and cyan ($27$ $R_p$). The vertical lines mark the initial semi-major by colour, the outermost vertical line marks $40$ $R_p$, the initial position of Kepler-1625 b-I. The shapes represent the initial eccentricity of the satellite $(e_0)$, $0.0$ (open circle), $0.1$ (letter "x"), $0.2$ (asterisk), $0.3$ (square), $0.4$ (filled circle), $0.5$ (upwards triangle), $0.6$ (downwards triangle), $0.7$ (diamond) and $0.8$ (pentagon). The bodies represented in the outermost vertical lines are the respective Kepler-1625 b-I for each case. The dashed dark-blue vertical line indicates the initial corotation radius of the system for the secondary satellite.}
\label{fig:a-e-final}
\end{center}
\end{figure*}
Statically, from the 160 different systems shown here, we found that exactly half of the systems are stable and half are unstable. From all the 80 unstable systems we found that the most common fate was ejection with 30 occurrences ($37.5$ $\%$), followed closely by collision with the primary satellite, 29 cases ($36.25$ $\%$), and collision with the central planet, 21 cases ($26.25$ $\%$). Since the most unstable systems were found closer to the primary satellite, the most common outcomes for instability being ejection and collision with this satellite are expected.

%% file: sec-04.tex
\section{Discussions}\label{sfour}
In this section, we will focus on the stable systems and discuss their dynamical evolution, such as migration, eccentricity damping and resonant configurations. Also, we will analyse the influences of the initial semi-major axis and eccentricity on our findings. \par  

\subsection{Migration and Eccentricity Damping}

In Fig. \ref{fig:a-e-final} we show the final distribution of semi-major axis vs eccentricity of our stable cases. When comparing the initial and final semi-major axes of the satellites, one can see that in most cases the satellites migrated inwards or were pushed inwards due to a close encounter with the primary satellite. Notably, we found several cases of stability inside $5$ $R_p$, a region where the gravitational and tidal effects of the planet are stronger. These satellites present low eccentricities and will be long-term stable once they are outside their respective corotation radius, which means there is no risk of these satellites colliding with the planet.\par
We found that the satellites migrating the most are the ones in initial eccentric orbits. Even though the secondary satellites are initially beyond the corotation radius of the systems, because of their eccentric orbits, we have the satellites tides prevailing over the planetary tides, in this case, as discussed before, the satellites will migrate inwards until their eccentricity is damped to values below $0.1$.\par
\begin{figure*}
\begin{center}
\includegraphics[height=0.75\linewidth, width=\linewidth]{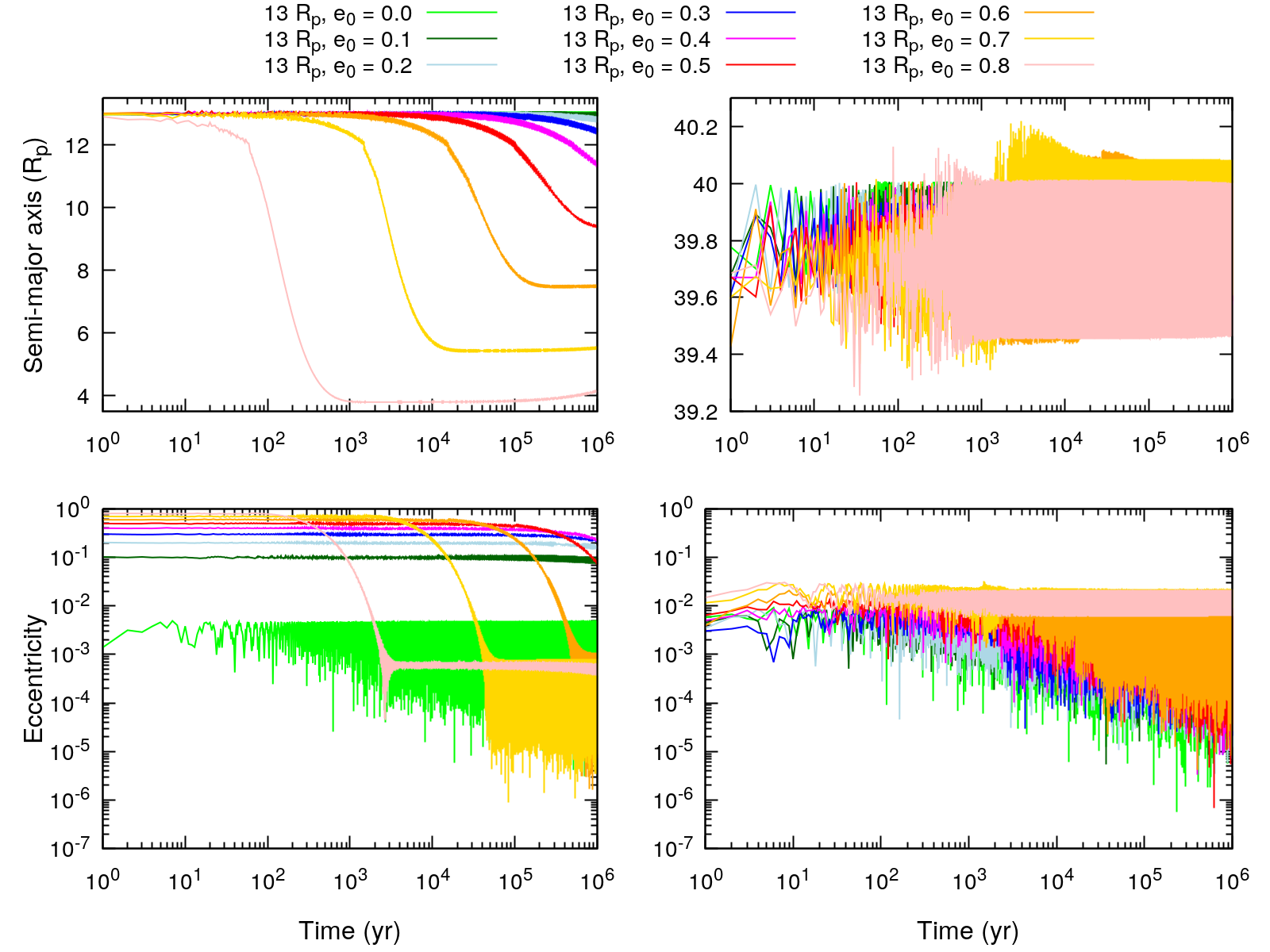}
\caption[13-Rp]{Time evolution of the semi-major axis and eccentricity of the primary and secondary satellites for the stable cases where the secondary satellites started at $13$ $R_p$. In the left panels, we have the evolution of the semi-major axis (upper panel) and eccentricity (lower panel) of the secondary satellites. In the right panels, we have the evolution of the semi-major axis (upper panel) and eccentricity (lower panel) of the primary satellites. The cases are separated by the initial eccentricity of the secondary satellites using a colour scheme, from $e_0=0.0$ to $e_0=0.8$ we have the following colours: light green, green, light blue, blue, magenta, red, orange, yellow and pink, respectively. The evolution of the case with $e_0=0.9$ is not shown because it resulted in an unstable system.}
\label{fig:13-Rp}
\end{center}
\end{figure*}
As an example, we show in Fig. \ref{fig:13-Rp} the evolution of the semi-major axis and eccentricity of the secondary (Earth-like) and the primary (Kepler-1625 b-I) satellites for the case where the Earth-like satellite started at $13$ $R_p$. The evolution of the semi-major axis and eccentricity of the secondary satellites are shown in the left panels, while the same parameters for the primary satellites are presented in the right panels. We differ each case by their initial eccentricity using a colour scheme, from $e_0=0.0$ to $e_0=0.8$ we have the following colours: light green, green, light blue, blue, magenta, red, orange, yellow and pink, respectively. The evolution of the case with $e_0=0.9$ is not displayed because it resulted in an unstable system. \par
As expected, the satellites with initially higher eccentricity migrated inwardly faster and reached inner orbits. In this phase, the migration of the eccentric satellites is being determined by the satellites tides regime. As the satellites migrate towards the planet their eccentricity is damped by the tidal interactions with the central body. As soon as the eccentricity of the satellites reached small values, lower than $0.1$ \citep{Bolmont-etal-2011}, the inward migration halts and at this point, the strength of the planetary tides overcome the satellite tides and the evolution of the semi-major axis of these bodies is determined by the planet-satellite spin ratio, which in the cases shown here result in a slow outwards migration. While the migration regime changed, the eccentricity of the satellites still decreases faster because of the proximity with the planet, eventually, this drop on the eccentricity stops and the eccentricity of the satellites reaches a non-zero equilibrium range of values. \par
It is worth mentioning the role of Kepler-1625 b-I on the evolution of the systems. From the right panel in Fig. \ref{fig:13-Rp} we see that the primary satellites' semi-major axis did not vary significantly, this is because these satellites are too far from the central planet to be tidally disturbed and their interaction with the inner satellite did not affect much their semi-major axis. However, the eccentricities of the primary satellites are excited by the presence of the inner satellites. On the other hand, the presence of the outer satellites is the reason for the inner satellites never having reached circular orbits, even with the tidal damping acting over their eccentricities. Also, for the case where the secondary satellite started in a circular orbit (light green line on the bottom left panel), we can see that the presence of Kepler-1625 b-I is crucial, such as the secondary satellite's orbit was perturbed and its eccentricity increased to values near $0.001$ almost instantaneously. Nevertheless, this non-zero eccentricity was not translated to a significant change on the semi-major axis of the Earth-like satellite in this specific case. \par
The pattern exemplified in Fig. \ref{fig:13-Rp} was found in most of our simulations, except the cases where the satellites were trapped in resonance or pushed out of a resonant configuration. Namely, all the cases where the secondary satellite had an initial semi-major axis smaller than $17$ $R_p$ had a similar outcome than the one showed before. For the systems with the secondary satellite initially in wider orbits, a combination of weaker tidal forces and the presence of low-order MMR prevent the significant inward migration of the secondary satellites. In these cases, most of the satellites did not migrate much and their final eccentricities are higher.\par

\subsection{Resonances}

\begin{figure*}
\begin{center}
\includegraphics[height=0.75\linewidth, width=\linewidth]{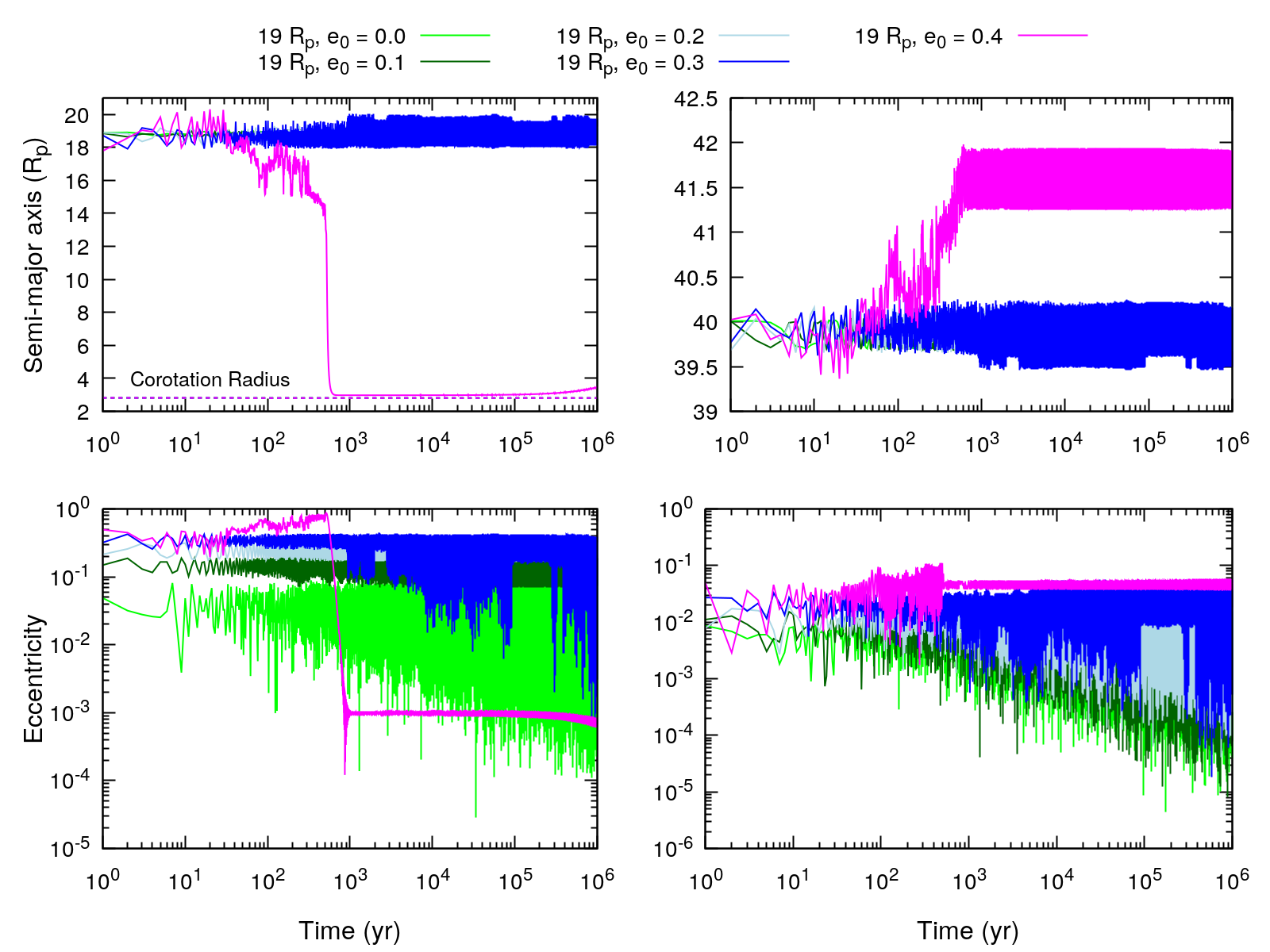}
\caption[19-Rp]{Time evolution of the semi-major axis and eccentricity of the primary and secondary satellites for the stable cases where the secondary satellites started at $19$ $R_p$ (near the 3:1 MMR). In the left panels, we have the evolution of the semi-major axis (upper panel) and eccentricity (lower panel) of the secondary satellites. In the right panels, we have the evolution of the semi-major axis (upper panel) and eccentricity (lower panel) of the primary satellites. The cases are separated by the initial eccentricity of the secondary satellites using a colour scheme, from $e_0=0.0$ to $e_0=0.4$ we have the following colours: light green, green, light blue, blue, magenta, respectively. The evolution of the other cases is not shown because they resulted in unstable systems. The horizontal dashed lines indicate the corotation radius.}
\label{fig:19-Rp}
\end{center}
\end{figure*}

In recent years compact multi-planetary systems locked in resonances were found \citep{Mills-etal-2016, Luger-etal-2017, Leleu-etal-2021}. According to \citet{Mills-etal-2016}, the presence of MMRs in compact systems greatly helps the stability of the systems exciting and stabilizing the eccentricity of the planets. In satellite systems, resonances also play a significant role not only in the stability of the system but to sculpt the final architecture of the system as well, the Laplace resonance presented by the Galilean system being the most known example in our Solar System.\par
Here we will discuss the role that a few MMRs play in the stability of the systems.\par

\subsubsection{Systems near the 3:1 MMR}
We start our analysis with systems that are not in resonance but very close. In Fig. \ref{fig:19-Rp} we show the evolution of the semi-major axis and eccentricity of the secondary and the primary satellites for the case where the Earth-like satellites started at $19$ $R_p$. As before, on the left side, we have the evolution of the semi-major axis and eccentricity of the secondary satellites, while on the right side, we have the same parameters presented for the primary satellites.\par
\begin{figure}
\begin{center}
\includegraphics[height=0.633\columnwidth, width=0.95\columnwidth]{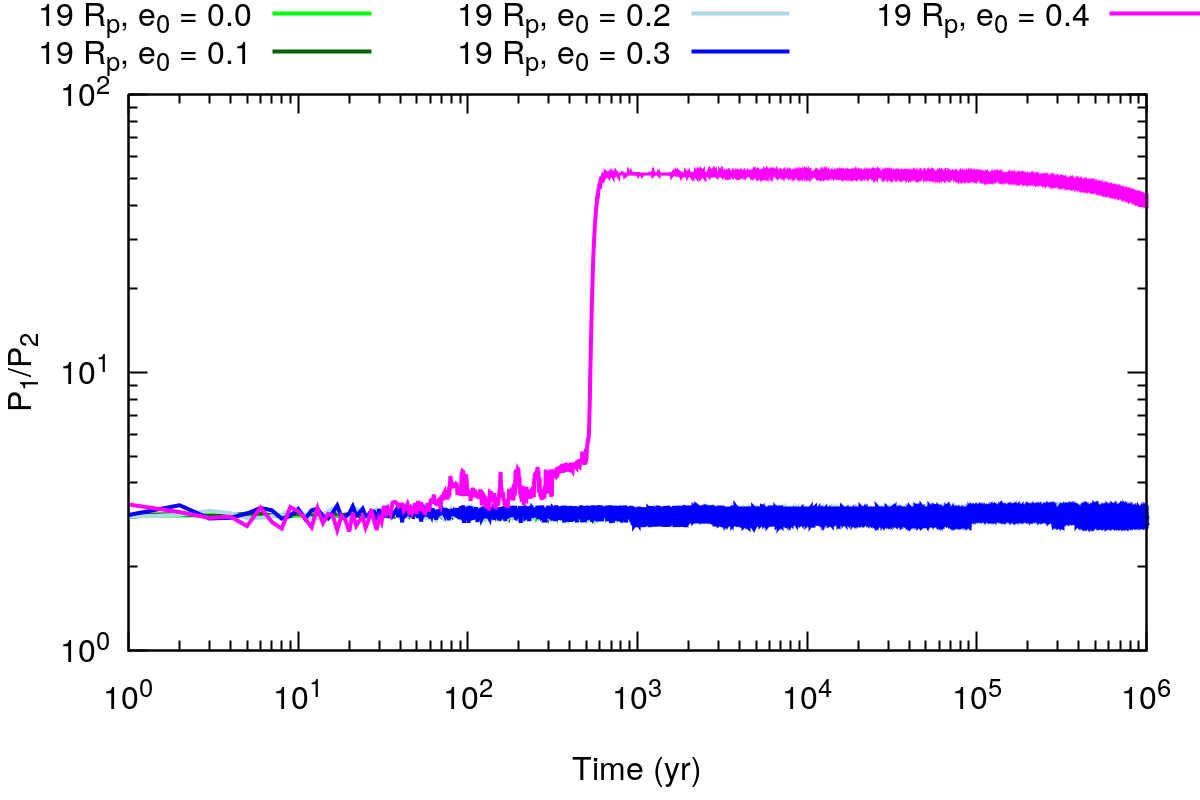}
\caption[Period-19-Rp]{Time evolution of the satellites' period ratio for the stable systems presented in Fig \ref{fig:19-Rp} (near the 3:1 MMR). }
\label{fig:period-19-Rp}
\end{center}
\end{figure}
\begin{figure*}
\begin{center}
\includegraphics[height=0.75\linewidth, width=\linewidth]{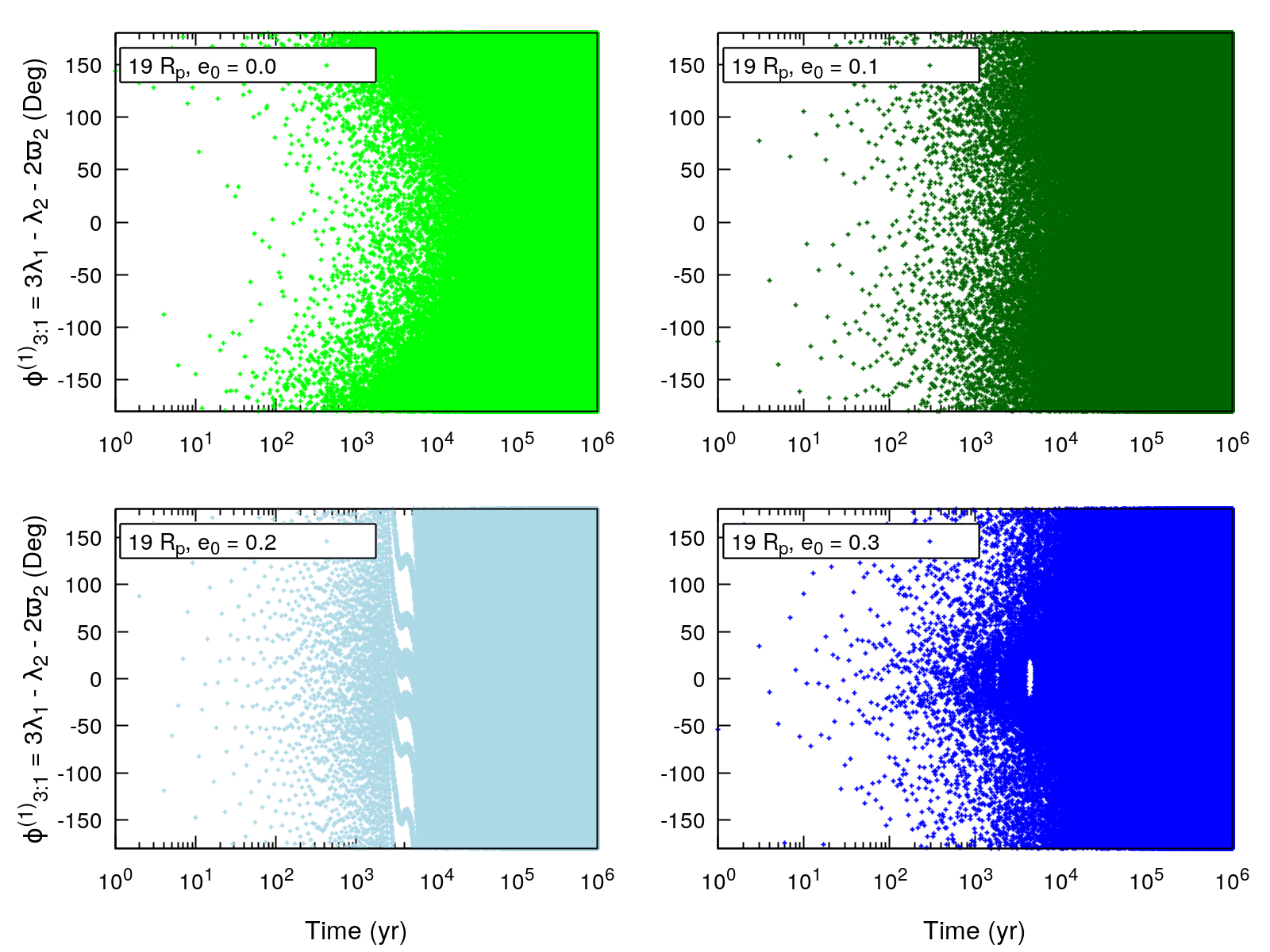}
\caption[19-Resonant-angle]{Time evolution of the resonant angle $\phi^{(1)}_{3:1} = 3\lambda_1 - \lambda_2 -2\varpi_2$ (first line in Eq. \ref{eq:phi-31}) for the stable cases where the secondary satellites started at $19$ $R_p$ (near the 3:1 MMR). From top left to bottom right we have the cases separated by the initial eccentricity of the secondary satellites, from $e_0=0.0$ to $e_0=0.3$ The case with $e_0 = 0.4$ was omitted from the graph because the system is clearly not in a MMR.}
\label{fig:19-resonant-angle}
\end{center}
\end{figure*}
At first, the evolution of the systems presented in Fig. \ref{fig:19-Rp} might resemble the evolution of systems locked in a MMR, where the motion of the secondary satellite is limited by the presence of the primary satellite while sustaining a non-zero eccentricity. In addition, in Fig. \ref{fig:period-19-Rp} we show the time evolution of the satellites' period ratio, $P_1/P_2$, where $P_1 = 2\pi/n_1$ is the orbital period of the primary satellite and $P_2 = 2\pi/n_2$ is the orbital period of the secondary satellite, with $n_1$ and $n_2$ being the respectively mean motion. The satellites' period ratio is approximately three at the beginning of our simulations and, except for one case, most of the satellites preserve this period commensurability until the end of their evolution, with a small amplitude. In this way, one could argue that the satellites might be locked in a 3:1 MMR. However, for this to be true we have to study the behaviour of the resonant angle of this hypothetical resonance. From \citet{Murray-1999} we have that the three resonant angles for a 3:1 MMR are
\begin{align}\label{eq:phi-31}
    \phi^{(1)}_{3:1} &= 3\lambda_1 - \lambda_2 -2\varpi_2,\\ \nonumber
    \phi^{(2)}_{3:1} &= 3\lambda_1 - \lambda_2 -2\varpi_1,\\ \nonumber
    \phi^{(3)}_{3:1} &= 3\lambda_1 - \lambda_2 -\varpi_1 - \varpi_2,
\end{align}
where $\lambda_1$ and $\varpi_1$ are the mean longitude and longitude of the pericenter for the primary satellite, respectively and $\lambda_2$ and $\varpi_2$ are the same quantities for the secondary satellite. \par
For two bodies in resonance, one of the three angles described in Eq. \ref{eq:phi-31} would be librating around one specific value. However, for all cases, the three angles were found to be circulating, which means that, despite the commensurability of the satellites' orbital period, the satellites are not locked in resonance. As an example of the time evolution of the resonant angles for these cases, we show in Fig. \ref{fig:19-resonant-angle} the behaviour of $\phi^{(1)}_{3:1}$ for the cases where the secondary satellites started with eccentricities from $e_0 = 0.0$ to $e_0 = 0.3$, the case with $e_0 = 0.4$ is not shown because the system is clearly out of a hypothetical MMR. The behaviours of $\phi^{(2)}_{3:1}$ and $\phi^{(3)}_{3:1}$ are similar to the one of $\phi^{(1)}_{3:1}$.\par
In this case, instead of a MMR locking the motion of the satellites, the semi-major axes and eccentricities of the satellites are not damped because at $19$ $R_p$ the tidal effects coming from the planet are weak and the satellites preserve an orbital configuration similar to their initial conditions. In this way, we cannot evaluate the role the 3:1 MMR plays for the stability of the satellites, since the satellites near this resonance did not show any resonant behaviour.\par
\begin{figure*}
\begin{center}
\includegraphics[height=0.75\linewidth, width=\linewidth]{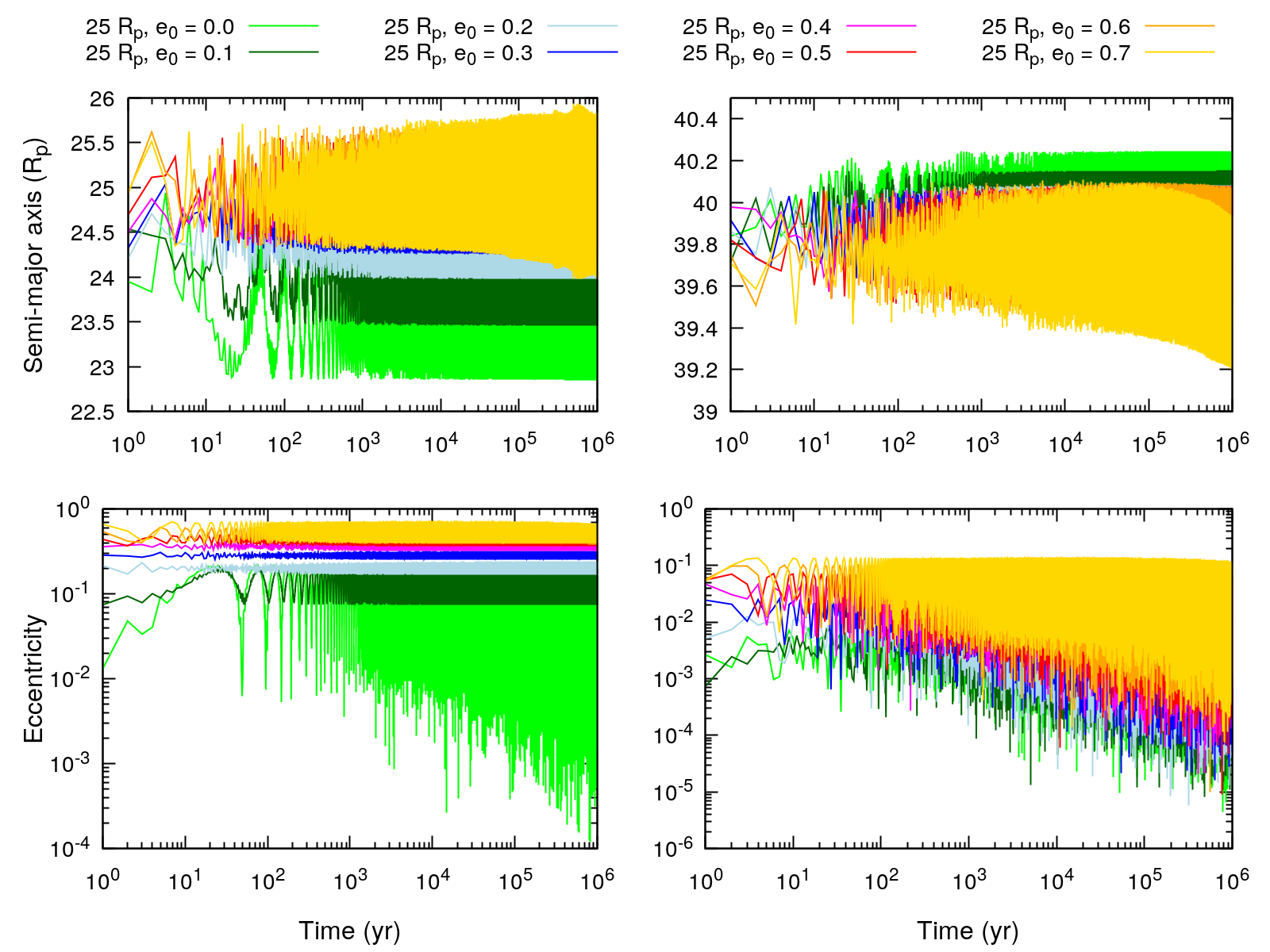}
\caption[25-Rp]{Time evolution of the semi-major axis and eccentricity of the primary and secondary satellites for the stable cases where the secondary satellites started at $25$ $R_p$ (near the 2:1 MMR). In the left panels, we have the evolution of the semi-major axis (upper panel) and eccentricity (lower panel) of the secondary satellites. In the right panels, we have the evolution of the semi-major axis (upper panel) and eccentricity (lower panel) of the primary satellites. The cases are separated by the initial eccentricity of the secondary satellites using a colour scheme, from $e_0=0.0$ to $e_0=0.7$ we have the following colours: light green, green, light blue, blue, magenta, red, orange and yellow, respectively. The evolution of the cases with $e_0=0.8$ and $e_0=0.9$ are not shown because they resulted in unstable systems.}
\label{fig:25-Rp}
\end{center}
\end{figure*}
Another important feature displayed in Fig. \ref{fig:19-Rp} is the behaviour of the secondary satellite that started with $e_0 = 0.4$ (magenta line). Different from the other cases, this satellite escaped the region near the 3:1 MMR thanks to multiple close encounters with the primary satellite, as can be seen by the erratic behaviour of its semi-major axis, before a final encounter that resulted in the secondary satellite being pushed inwards. Because the damping effects of the tides over the satellites' eccentricity are stronger in the inner regions, the satellite's orbit was almost circularized before the satellite reaches the corotation radius. After this abrupt process of circularization, the satellite started to experiment an outward migration, resulting from the tides generated by the planet.\par
Since the secondary satellites around $19$ $R_p$ are not in MMR they are not free from a future close encounter with the primary satellite. As these satellites maintained high eccentricities during their evolution, a close encounter could take place at some point.

\begin{figure*}
\begin{center}
\includegraphics[height=0.888\linewidth, width=\linewidth]{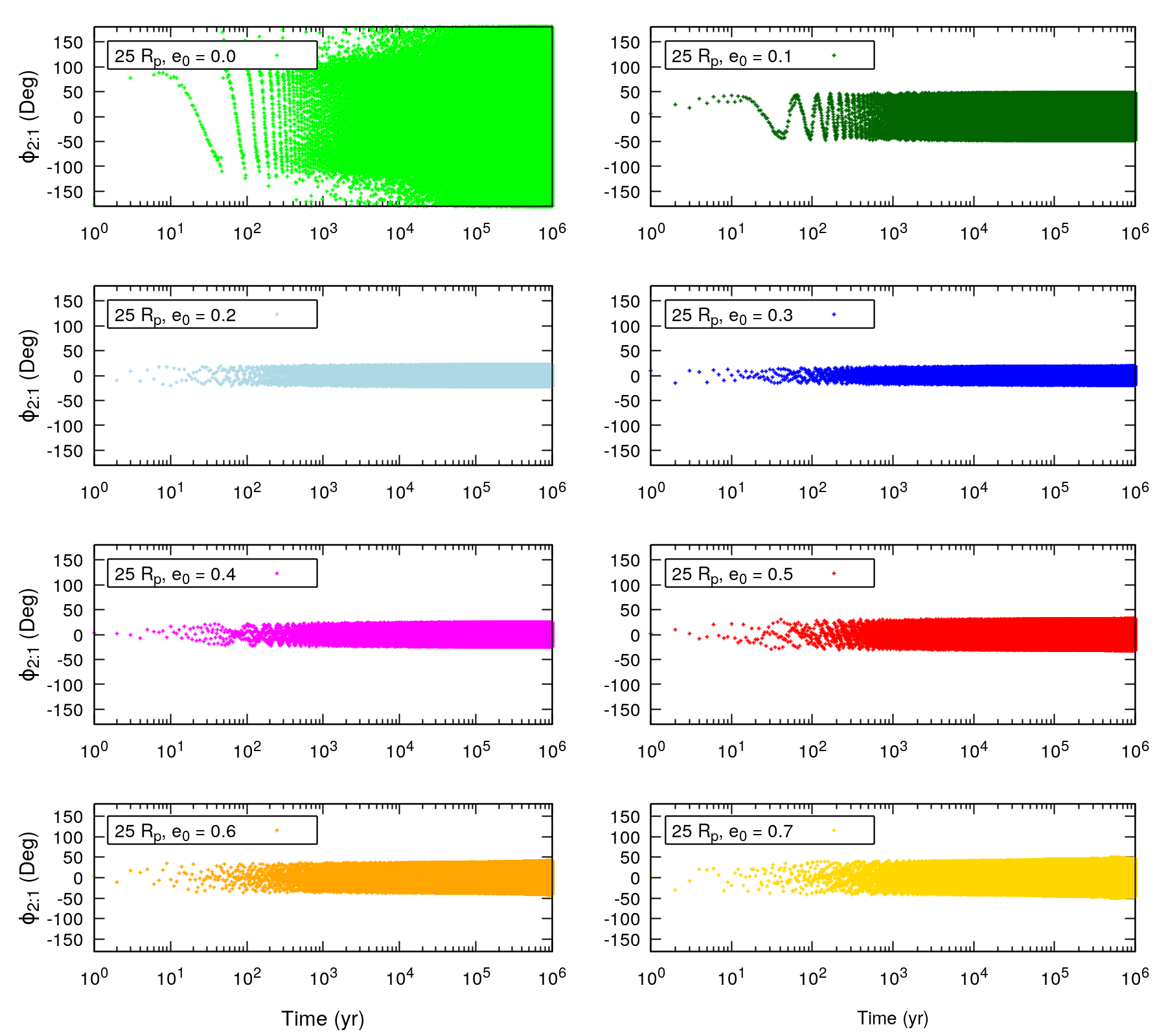}
\caption[25-Resonant-angle]{Time evolution of the resonant angle $\phi_{2:1} = 2\lambda_1 - \lambda_2 -\varpi_2$ for the stable cases where the secondary satellites started at $25$ $R_p$ (near the 2:1 MMR). From top left to bottom right we have the cases separated by the initial eccentricity of the secondary satellites, from $e_0=0.0$ to $e_0=0.7$.}
\label{fig:25-resonant-angle}
\end{center}
\end{figure*}

\subsubsection{Systems locked in 2:1 MMR}
As mentioned before, as we started the secondary satellites farther from the planet the effects of the tides become weaker and the mutual gravitational interaction between the satellites are stronger. In this way, we found that when the secondary satellites started beyond $20$ $R_p$ the percentage of stable systems decreased steeply. At these outer locations, the migration rate and eccentricity damping induced by the tides are smooth and the secondary satellites are more likely to suffer a disruptive close encounter with the primary satellite, especially the satellites in eccentric orbits. However, for the cases where the secondary satellites started at $25$ $R_p$ a high number of systems were stable, this is due to the presence of a 2:1 MMR near this initial position.\par
In Fig. \ref{fig:25-Rp} we show the evolution of the semi-major axis and eccentricity of the primary and secondary satellites for the stable system when the secondary satellite started at $25$ $R_p$. The high number of stable systems presented might be explained by the location of a 2:1 MMR with the primary satellite near the orbital distance where the satellites were initialized. Also, from Fig. \ref{fig:25-Rp} one can see that, despite small amplitudes, after 100 years the semi-major axis of the secondary satellites did not change significantly, the same is true for the eccentricities of the satellites that remain high in some cases.\par
Similar to the cases with the secondary satellites starting at $19$ $R_p$ it seems that the systems presented in Fig. \ref{fig:25-Rp} display a resonant configuration between the satellites. To validate this hypothesis once again, we will study the resonant angle for each system, in this case, the resonant angle refers to a 2:1 MMR and it is given by,
\begin{align}\label{eq:phi-21}
    \phi_{2:1} = 2\lambda_1 - \lambda_2 -\varpi_2.
\end{align}
In Fig. \ref{fig:25-resonant-angle} we show the resonant angle for all the stable systems with the secondary satellite starting at $25$ $R_p$.\par
The analysis of Fig. \ref{fig:25-resonant-angle} is straightforward. The case where the Earth-like satellite was initialized in circular orbit presents a circulating resonant angle, which means that the satellites are not locked in a MMR. From Fig. \ref{fig:25-Rp} one can see that this case is the one where the semi-major axis deviates the most from its initial value. The secondary satellite started at $25$ $R_p$, had a close encounter with the primary satellite (which is clear given the increase in its eccentricity and sudden decrease in semi-major axis soon after 10 years) and found a stable, but not resonant, configuration around $23$ $R_p$.\par
Except for the case where the Earth-like satellite was initialized in a circular orbit, all other cases present a resonant angle librating around $0^{\circ}$ with a maximum amplitude of $50^{\circ}$. This behaviour supports the claim that the satellites have their motion locked in a 2:1 MMR. Since the secondary satellites placed initially farther from the planet tend to become unstable with time, we argue that the 2:1 MMR between the satellites can be the dynamical mechanism assuring the stability of these systems. On the other hand, in some resonant systems the secondary satellite is in a very eccentric orbit, which might lead to a disruptive close encounter and the break of the resonant configuration between the satellites, such a dynamical event could jeopardize the stability of the systems.\par
\subsubsection{Unstable systems near the 3:2 MMR}
Beyond $25$ $R_p$ we have the presence of a 3:2 MMR resonance located near $31$ $R_p$. However, this region was found to be empty of stable systems. The systems with the secondary satellite starting at $31$ $R_p$ became unstable almost instantaneously given the strong gravitational influence of the primary satellite. For these systems, the most common fate for the secondary satellite was a collision with the primary satellite ($50\%$ of the systems), followed by ejection ($30\%$ of the systems) and collision with the planet ($20\%$).\par
Another reason for the lack of stable systems in the 3:2 MMR location could be the overlapping of resonances. As the distances from the primary satellite are shortened, successive first-order resonances of the form $p+1:p$ will appear, each one independent of the other. As each resonance has its width, at short distances from the exterior perturber these resonances will eventually start to overlap. As a consequence of this effect, the chaotic motion could arise leading to instability.\par
The location where resonance overlapping is preset can be estimated analytically in some simpler cases, for example, the planar circular-restricted three-body problem \citep{Wisdom-1980}. For a circular-restricted three-body problem where the particle has small eccentricity $(e\le 0.15)$, \citet{Murray-1999} estimated the half-width from the perturber where the resonances will start to overlap as,
\begin{align}\label{eq:overlap}
    \Delta a_{overlap} \approx 1.24 \mu^{2/7} a_{out},
\end{align}
where $\mu = M_{out}/(M_{in} + M_{out})$ is the mass of the external perturber weighted by the sum of the masses of the central body and the perturber respectively, and $a_{out}$ is the semi-major axis of the external perturber. Particles at $a_{out}\pm \Delta_{overlap}$ will present chaotic orbits. Extrapolating Eq. \ref{eq:overlap} for our systems, assuming $M_{in} = M_p = 3$ $M_{Jup}$, $M_{out} = M_1 = 1$ $M_{Nep}$ and $a_{out} = a_{1} = 40$ $R_p$ we have $\Delta_{overlap}\approx 15.6$ $R_p$, which means that the whole region outside $\approx 24.4$ $R_p$ would be chaotic. This is only a rough estimation for our cases because the secondary satellite is not a massless particle and its only valid for the satellites starting with $e_0<0.2$, since Eq. \ref{eq:overlap} is not applicable for cases where $e > 0.15$. However, this approximation somewhat fits our results, as we found that just a single initial condition starting beyond $25$ $R_p$ resulted in a stable system.\par
We point out that the regions of instability will change if the masses of the bodies considered and the initial radial separation of the primary satellite were different.

%% file: conclusion.tex
\section{Conclusions}
\label{conclusion} 
In this work, we have explored the dynamical stability of an Earth-like satellite in the Kepler-1625 b system, considering that the system already has a massive satellite, the Neptunian-like candidate Kepler-1625 b-I.\par
In this study, we tried to answer the following questions: 
\begin{enumerate}
    \item Given that Kepler-1625 b-I is stable, is it possible to have another massive moon in this system?
    \item Are there preferred orbital configurations or locations for the extra satellite?
    \item How Kepler-1625 b-I will dynamically affect this extra satellite? 
\end{enumerate}
To answer these questions, we performed N-body simulations considering a system formed by the central planet, Kepler-1625 b-I and an extra Earth-sized satellite using the N-body code POSIDONIUS. To accurately model the system, we took into account the tidal effects generated by the satellites on the planet and by the planet on the satellites, and the effects due to the rotation deformation of the simulated bodies.\par
We focus on the region between the planet and Kepler-1625 b-I (initially at $40$ $R_p$), such as we build a grid of initial conditions varying the semi-major axis and the eccentricity of the satellites. For the semi-major axis of the secondary satellites we chose to place the innermost case at $a_0 = 5$ $R_p$, outside the corotation radius of the system, while $a_0 = 35$ $R_p$ was taken as the outermost initial semi-major axis for the secondary satellites, using $\Delta a_0 = 2$ $R_p$ as the width of our grid. For each initial semi-major axis we simulated ten different cases varying the eccentricity of the secondary satellite, from $e_0 = 0.0$ to $e_0 = 0.9$. We opted to include highly eccentric cases in our grid because we are considering the tidal effects over the satellites, which could damp the eccentricity of such bodies to values close to zero. Thus, we have in total 160 different initial conditions.\par
In Fig. \ref{fig:a-e-results} we already answer our first question. Yes, it is possible to have two massive satellites, an Earth-like and a Neptune-like satellite, around the planet Kepler-1625 b. We found that $50\%$ of our simulations ended with a system with two satellites stable for $1$ Myr. As expected, the satellites with initial high eccentricity and orbits closer to the planet suffered a collision with the central body almost instantaneously. The same was true for the satellites initially in wider orbits, closer to Kepler-1625 b-I. Our results show that the cases where the secondary satellites were initially within $20$ $R_p$ have a higher chance to produce stable systems, even for initially eccentric configurations. \par
The second main question we try to answer is related to the final location of the surviving secondary satellites. We found that the primary satellite tends to remain at $\sim 40$ $R_p$ for most of the cases, unless close encounters between the satellites takes place, while its orbit acquired some non-negligible eccentricity (between $0.001$ and $0.1$) due to the gravitational interaction with the inner satellites.\par
For the Earth-like satellites, our results showed that there is no preferred location for these satellites to be stable inside $25$ $R_p$, beyond this limit the satellite will be mostly unstable due to gravitational effects from the primary satellite generating chaotic motion produced by resonance overlapping. For the surviving systems, we found that most of the secondary satellites migrated inwards, especially the eccentric satellites (due to the action of the satellite tides)\par
In addition to the satellites that experienced inward migration, we have satellites that for one reason or another did not present a significant migration. These satellites are either in circular orbits, too far to experience strong tidal interactions with the planet or trapped in MMRs with the primary satellite. \par
The third main question we answer is related to the dynamical effects of Kepler-1625 b-I over an inner massive satellite. As mentioned before, most of the satellites initially placed closer to the primary satellite $(a_0 > 25$ $R_p)$ became unstable. \par
Our results show that MMRs are responsible for most of the stable systems when the secondary satellite started at wider orbits. We kept track of satellites starting close to the 2:1, 3:2 and 3:1 MMRs. Only the 2:1 MMR effectively influenced the stability of the satellites. The 3:2 resonance was located too close to the primary satellite, thus the satellites near this MMR were unstable. On the other hand, the stable satellites near the 3:1 commensurability never became locked in resonance and this resonance did not protect the most eccentric satellites from being unstable.\par
In summary, our work shows that the satellite system around Kepler-1625 b can be even more fascinating than it is currently predicted. The system has the potential to harbour not only one, but two planet-sized satellites. This theoretical secondary satellite does not show a preferable location inside $25$ $R_p$ and could be stable in almost circular orbits close to the planet or in wider orbits locked in a MMR with Kepler-1625 b-I.\par
In this work, we restrict ourselves to studying the stability of regions between the planet and the predicted location of Kepler-1625 b-I, leaving the study of regions beyond $40$ $R_p$ and co-orbital configurations to further analysis. These two studies will be complementary to the one we present here since a more extended region of the system will be explored. \par
Also, we only study the coplanar case, assuming that all bodies involved had zero inclination. However, Kepler-1625 b-I, if confirmed, it is predicted to have a high inclination. In this case, an additional study taken into account inclined bodies is needed. We expect that the addition of inclined bodies will not affect the stability of the innermost surviving satellites once their evolution will be still dictated by the tides from the planet. On the other hand, the outer satellites will be more likely to be affected by an inclined body and eventually become inclined or unstable. Moreover, the stability of the outermost satellites rely on the appearance of MMRs, this could be an issue when inclined bodies are considered.

%% file: paper.bbl
\begin{thebibliography}{}
\makeatletter
\relax
\def\mn@urlcharsother{\let\do\@makeother \do\$\do\&\do\#\do\^\do\_\do\%\do\~}
\def\mn@doi{\begingroup\mn@urlcharsother \@ifnextchar [ {\mn@doi@}
  {\mn@doi@[]}}
\def\mn@doi@[#1]#2{\def\@tempa{#1}\ifx\@tempa\@empty \href
  {http://dx.doi.org/#2} {doi:#2}\else \href {http://dx.doi.org/#2} {#1}\fi
  \endgroup}
\def\mn@eprint#1#2{\mn@eprint@#1:#2::\@nil}
\def\mn@eprint@arXiv#1{\href {http://arxiv.org/abs/#1} {{\tt arXiv:#1}}}
\def\mn@eprint@dblp#1{\href {http://dblp.uni-trier.de/rec/bibtex/#1.xml}
  {dblp:#1}}
\def\mn@eprint@#1:#2:#3:#4\@nil{\def\@tempa {#1}\def\@tempb {#2}\def\@tempc
  {#3}\ifx \@tempc \@empty \let \@tempc \@tempb \let \@tempb \@tempa \fi \ifx
  \@tempb \@empty \def\@tempb {arXiv}\fi \@ifundefined
  {mn@eprint@\@tempb}{\@tempb:\@tempc}{\expandafter \expandafter \csname
  mn@eprint@\@tempb\endcsname \expandafter{\@tempc}}}

\bibitem[\protect\citeauthoryear{{Agnor} \& {Hamilton}}{{Agnor} \&
  {Hamilton}}{2006}]{Agnor-Hamilton-2006}
{Agnor} C.~B.,  {Hamilton} D.~P.,  2006, \mn@doi [Nature]
  {10.1038/nature04792}, \href
  {https://ui.adsabs.harvard.edu/abs/2006Natur.441..192A} {441, 192}

\bibitem[\protect\citeauthoryear{{\'A}vila, Grassi, Bovino, Chiavassa,
  Ercolano, Danielache  \& Simoncini}{{\'A}vila et~al.}{2021}]{Avila-etal-2021}
{\'A}vila P.~J.,  Grassi T.,  Bovino S.,  Chiavassa A.,  Ercolano B.,
  Danielache S.~O.,   Simoncini E.,  2021, \mn@doi [International Journal of
  Astrobiology] {10.1017/S1473550421000173}, pp 1--12

\bibitem[\protect\citeauthoryear{{Barr}}{{Barr}}{2016}]{Barr-2016}
{Barr} A.~C.,  2016, \mn@doi [The Astronomical Review]
  {10.1080/21672857.2017.1279469}, \href
  {http://adsabs.harvard.edu/abs/2016AstRv.12...24B} {12, 24}

\bibitem[\protect\citeauthoryear{{Ben-Jaffel} \& {Ballester}}{{Ben-Jaffel} \&
  {Ballester}}{2014}]{Ben-Jaffel-Ballester-2014}
{Ben-Jaffel} L.,  {Ballester} G.~E.,  2014, \mn@doi [ApJL]
  {10.1088/2041-8205/785/2/L30}, \href
  {http://adsabs.harvard.edu/abs/2014ApJ...785L..30B} {785, L30}

\bibitem[\protect\citeauthoryear{{Bennett} et~al.,}{{Bennett}
  et~al.}{2014}]{Bennett-etal-2014}
{Bennett} D.~P.,  et~al., 2014, \mn@doi [ApJ] {10.1088/0004-637X/785/2/155},
  \href {http://adsabs.harvard.edu/abs/2014ApJ...785..155B} {785, 155}

\bibitem[\protect\citeauthoryear{Blanco-Cuaresma \& Bolmont}{Blanco-Cuaresma \&
  Bolmont}{2016}]{Blanco-Cuaresma-Bolmont-2017b}
Blanco-Cuaresma S.,  Bolmont E.,  2016. Cambridge University Press, pp
  341--344, \mn@doi{10.1017/S1743921316013168}

\bibitem[\protect\citeauthoryear{{Blanco-Cuaresma} \&
  {Bolmont}}{{Blanco-Cuaresma} \&
  {Bolmont}}{2017}]{Blanco-Cuaresma-Bolmont-2017a}
{Blanco-Cuaresma} S.,  {Bolmont} E.,  2017, in EWASS Special Session 4 (2017):
  Star-planet interactions (EWASS-SS4-2017).  (\mn@eprint {arXiv}
  {1712.01281}), \mn@doi{10.5281/zenodo.1095095}

\bibitem[\protect\citeauthoryear{{Bolmont}, {Raymond}  \& {Leconte}}{{Bolmont}
  et~al.}{2011}]{Bolmont-etal-2011}
{Bolmont} E.,  {Raymond} S.~N.,   {Leconte} J.,  2011, \mn@doi [A \& A]
  {10.1051/0004-6361/201117734}, \href
  {https://ui.adsabs.harvard.edu/abs/2011A&A...535A..94B} {535, A94}

\bibitem[\protect\citeauthoryear{{Bolmont}, {Selsis}, {Raymond}, {Leconte},
  {Hersant}, {Maurin}  \& {Pericaud}}{{Bolmont}
  et~al.}{2013}]{Bolmont-etal-2013}
{Bolmont} E.,  {Selsis} F.,  {Raymond} S.~N.,  {Leconte} J.,  {Hersant} F.,
  {Maurin} A.-S.,   {Pericaud} J.,  2013, \mn@doi [A \& A]
  {10.1051/0004-6361/201220837}, \href
  {https://ui.adsabs.harvard.edu/abs/2013A&A...556A..17B} {556, A17}

\bibitem[\protect\citeauthoryear{{Bolmont}, {Raymond}, {Leconte}, {Hersant}  \&
  {Correia}}{{Bolmont} et~al.}{2015}]{Bolmont-etal-2015}
{Bolmont} E.,  {Raymond} S.~N.,  {Leconte} J.,  {Hersant} F.,   {Correia} A.
  C.~M.,  2015, \mn@doi [A \& A] {10.1051/0004-6361/201525909}, \href
  {https://ui.adsabs.harvard.edu/abs/2015A&A...583A.116B} {583, A116}

\bibitem[\protect\citeauthoryear{{Bolmont}, {Oza}, {Blanco-Cuaresma},
  {Mordasini}, {Auclair-Desrotour}  \& {Leleu}}{{Bolmont}
  et~al.}{2019}]{Bolmont-etal-2019}
{Bolmont} E.,  {Oza} A.,  {Blanco-Cuaresma} S.,  {Mordasini} C.,
  {Auclair-Desrotour} P.,   {Leleu} A.,  2019, in EPSC-DPS Joint Meeting 2019.
  pp EPSC--DPS2019--1590

\bibitem[\protect\citeauthoryear{{Bolmont}, {Demory}, {Blanco-Cuaresma},
  {Agol}, {Grimm}, {Auclair-Desrotour}, {Selsis}  \& {Leleu}}{{Bolmont}
  et~al.}{2020}]{Bolmont-etal-2020}
{Bolmont} E.,  {Demory} B.~O.,  {Blanco-Cuaresma} S.,  {Agol} E.,  {Grimm}
  S.~L.,  {Auclair-Desrotour} P.,  {Selsis} F.,   {Leleu} A.,  2020, \mn@doi [A
  \& A] {10.1051/0004-6361/202037546}, \href
  {https://ui.adsabs.harvard.edu/abs/2020A&A...635A.117B} {635, A117}

\bibitem[\protect\citeauthoryear{Correia \& Rodr{\'{\i}}guez}{Correia \&
  Rodr{\'{\i}}guez}{2013}]{Correia-Rodriguez-2013}
Correia A. C.~M.,  Rodr{\'{\i}}guez A.,  2013, \mn@doi [The Astrophysical
  Journal] {10.1088/0004-637x/767/2/128}, 767, 128

\bibitem[\protect\citeauthoryear{{Fox} \& {Wiegert}}{{Fox} \&
  {Wiegert}}{2021}]{Fox-Wiegert-2021}
{Fox} C.,  {Wiegert} P.,  2021, \mn@doi [MNRAS] {10.1093/mnras/staa3743}, \href
  {https://ui.adsabs.harvard.edu/abs/2021MNRAS.501.2378F} {501, 2378}

\bibitem[\protect\citeauthoryear{{Hamers} \& {Portegies Zwart}}{{Hamers} \&
  {Portegies Zwart}}{2018}]{Hamers-Zwart-2018}
{Hamers} A.~S.,  {Portegies Zwart} S.~F.,  2018, \mn@doi [ApJL]
  {10.3847/2041-8213/aaf3a7}, \href
  {http://adsabs.harvard.edu/abs/2018ApJ...869L..27H} {869, L27}

\bibitem[\protect\citeauthoryear{{Hansen}}{{Hansen}}{2010}]{Hansen-2010}
{Hansen} B. M.~S.,  2010, \mn@doi [ApJ] {10.1088/0004-637X/723/1/285}, \href
  {https://ui.adsabs.harvard.edu/abs/2010ApJ...723..285H} {723, 285}

\bibitem[\protect\citeauthoryear{{Hansen}}{{Hansen}}{2019}]{Hansen-2019}
{Hansen} B. M.~S.,  2019, \mn@doi [Science Advances] {10.1126/sciadv.aaw8665},
  \href {https://ui.adsabs.harvard.edu/abs/2019SciA....5.8665H} {5, eaaw8665}

\bibitem[\protect\citeauthoryear{{Heller}}{{Heller}}{2018}]{Heller-2018}
{Heller} R.,  2018, \mn@doi [A \& A] {10.1051/0004-6361/201731760}, \href
  {http://adsabs.harvard.edu/abs/2018A2%26A...610A..39H} {610, A39}

\bibitem[\protect\citeauthoryear{{Heller} \& {Pudritz}}{{Heller} \&
  {Pudritz}}{2015a}]{Heller-Pudritz-2015a}
{Heller} R.,  {Pudritz} R.,  2015a, \mn@doi [A \& A]
  {10.1051/0004-6361/201425487}, \href
  {http://adsabs.harvard.edu/abs/2015A26A...578A..19H} {578, A19}

\bibitem[\protect\citeauthoryear{{Heller} \& {Pudritz}}{{Heller} \&
  {Pudritz}}{2015b}]{Heller-Pudritz-2015b}
{Heller} R.,  {Pudritz} R.,  2015b, \mn@doi [ApJ]
  {10.1088/0004-637X/806/2/181}, \href
  {http://adsabs.harvard.edu/abs/2015ApJ...806..181H} {806, 181}

\bibitem[\protect\citeauthoryear{{Heller} et~al.,}{{Heller}
  et~al.}{2014}]{Heller-etal-2014}
{Heller} R.,  et~al., 2014, \mn@doi [Astrobiology] {10.1089/ast.2014.1147},
  \href {http://adsabs.harvard.edu/abs/2014AsBio..14..798H} {14, 798}

\bibitem[\protect\citeauthoryear{{Heller}, {Hippke}, {Placek}, {Angerhausen}
  \& {Agol}}{{Heller} et~al.}{2016}]{Heller-etal-2016}
{Heller} R.,  {Hippke} M.,  {Placek} B.,  {Angerhausen} D.,   {Agol} E.,  2016,
  \mn@doi [A \& A] {10.1051/0004-6361/201628573}, \href
  {https://ui.adsabs.harvard.edu/abs/2016A&A...591A..67H} {591, A67}

\bibitem[\protect\citeauthoryear{{Heller}, {Rodenbeck}  \& {Bruno}}{{Heller}
  et~al.}{2019}]{Heller-etal-2019}
{Heller} R.,  {Rodenbeck} K.,   {Bruno} G.,  2019, \mn@doi [A \& A]
  {10.1051/0004-6361/201834913}, \href
  {https://ui.adsabs.harvard.edu/abs/2019A&A...624A..95H} {624, A95}

\bibitem[\protect\citeauthoryear{{Hippke}}{{Hippke}}{2015}]{Hippke-2015}
{Hippke} M.,  2015, \mn@doi [ApJ] {10.1088/0004-637X/806/1/51}, \href
  {http://adsabs.harvard.edu/abs/2015ApJ...806...51H} {806, 51}

\bibitem[\protect\citeauthoryear{{Hussmann}, {Rodr{\'\i}guez}, {Callegari}  \&
  {Shoji}}{{Hussmann} et~al.}{2019}]{Hussmann-etal-2019}
{Hussmann} H.,  {Rodr{\'\i}guez} A.,  {Callegari} N.,   {Shoji} D.,  2019,
  \mn@doi [Icarus] {10.1016/j.icarus.2018.09.025}, \href
  {https://ui.adsabs.harvard.edu/abs/2019Icar..319..407H} {319, 407}

\bibitem[\protect\citeauthoryear{{Hut}}{{Hut}}{1981}]{Hut-1981}
{Hut} P.,  1981, A \& A, \href
  {https://ui.adsabs.harvard.edu/abs/1981A&A....99..126H} {99, 126}

\bibitem[\protect\citeauthoryear{{Kipping}}{{Kipping}}{2021}]{Kipping-2021}
{Kipping} D.,  2021, \mn@doi [MNRAS] {10.1093/mnras/staa3398}, \href
  {https://ui.adsabs.harvard.edu/abs/2021MNRAS.500.1851K} {500, 1851}

\bibitem[\protect\citeauthoryear{{Kollmeier} \& {Raymond}}{{Kollmeier} \&
  {Raymond}}{2019}]{Kollmeier-Raymond-2019}
{Kollmeier} J.~A.,  {Raymond} S.~N.,  2019, \mn@doi [MNRAS]
  {10.1093/mnrasl/sly219}, \href
  {https://ui.adsabs.harvard.edu/abs/2019MNRAS.483L..80K} {483, L80}

\bibitem[\protect\citeauthoryear{{Kreidberg}, {Luger}  \& {Bedell}}{{Kreidberg}
  et~al.}{2019}]{Kreidberg-etal-2019}
{Kreidberg} L.,  {Luger} R.,   {Bedell} M.,  2019, \mn@doi [ApJL]
  {10.3847/2041-8213/ab20c8}, \href
  {https://ui.adsabs.harvard.edu/abs/2019ApJ...877L..15K} {877, L15}

\bibitem[\protect\citeauthoryear{{Leconte}, {Chabrier}, {Baraffe}  \&
  {Levrard}}{{Leconte} et~al.}{2010}]{Leconte-etal-2010}
{Leconte} J.,  {Chabrier} G.,  {Baraffe} I.,   {Levrard} B.,  2010, \mn@doi [A
  \& A] {10.1051/0004-6361/201014337}, \href
  {https://ui.adsabs.harvard.edu/abs/2010A&A...516A..64L} {516, A64}

\bibitem[\protect\citeauthoryear{{Leleu} et~al.,}{{Leleu}
  et~al.}{2021}]{Leleu-etal-2021}
{Leleu} A.,  et~al., 2021, \mn@doi [A \& A] {10.1051/0004-6361/202039767},
  \href {https://ui.adsabs.harvard.edu/abs/2021A&A...649A..26L} {649, A26}

\bibitem[\protect\citeauthoryear{{Lewis}, {Ochiai}, {Nagasawa}  \&
  {Ida}}{{Lewis} et~al.}{2015}]{Lewis-etal-2015}
{Lewis} K.~M.,  {Ochiai} H.,  {Nagasawa} M.,   {Ida} S.,  2015, \mn@doi [ApJ]
  {10.1088/0004-637X/805/1/27}, \href
  {http://adsabs.harvard.edu/abs/2015ApJ...805...27L} {805, 27}

\bibitem[\protect\citeauthoryear{{Luger} et~al.,}{{Luger}
  et~al.}{2017}]{Luger-etal-2017}
{Luger} R.,  et~al., 2017, \mn@doi [Nature Astronomy]
  {10.1038/s41550-017-0129}, \href
  {http://adsabs.harvard.edu/abs/2017NatAs...1E.129L} {1, 1}

\bibitem[\protect\citeauthoryear{Martin, Fabrycky  \& Montet}{Martin
  et~al.}{2019}]{Martin-etal-2019}
Martin D.~V.,  Fabrycky D.~C.,   Montet B.~T.,  2019, \mn@doi [ApJL]
  {10.3847/2041-8213/ab0aea}, 875, L25

\bibitem[\protect\citeauthoryear{{Mathur} et~al.,}{{Mathur}
  et~al.}{2017}]{Mathur-etal-2017}
{Mathur} S.,  et~al., 2017, \mn@doi [ApJS] {10.3847/1538-4365/229/2/30}, \href
  {http://adsabs.harvard.edu/abs/2017ApJS..229...30M} {229, 30}

\bibitem[\protect\citeauthoryear{{Mignard}}{{Mignard}}{1979}]{Mignard-1979}
{Mignard} F.,  1979, \mn@doi [Moon and Planets] {10.1007/BF00907581}, \href
  {https://ui.adsabs.harvard.edu/abs/1979M&P....20..301M} {20, 301}

\bibitem[\protect\citeauthoryear{{Mills}, {Fabrycky}, {Migaszewski}, {Ford},
  {Petigura}  \& {Isaacson}}{{Mills} et~al.}{2016}]{Mills-etal-2016}
{Mills} S.~M.,  {Fabrycky} D.~C.,  {Migaszewski} C.,  {Ford} E.~B.,  {Petigura}
  E.,   {Isaacson} H.,  2016, \mn@doi [Nature] {10.1038/nature17445}, \href
  {https://ui.adsabs.harvard.edu/abs/2016Natur.533..509M} {533, 509}

\bibitem[\protect\citeauthoryear{{Moraes} \& {Vieira Neto}}{{Moraes} \& {Vieira
  Neto}}{2020}]{Moraes-Vieira-Neto-2020}
{Moraes} R.~A.,  {Vieira Neto} E.,  2020, \mn@doi [MNRAS]
  {10.1093/mnras/staa1441}, \href
  {https://ui.adsabs.harvard.edu/abs/2020MNRAS.495.3763M} {495, 3763}

\bibitem[\protect\citeauthoryear{{Morton}, {Bryson}, {Coughlin}, {Rowe},
  {Ravichandran}, {Petigura}, {Haas}  \& {Batalha}}{{Morton}
  et~al.}{2016}]{Morton-etal-2016}
{Morton} T.~D.,  {Bryson} S.~T.,  {Coughlin} J.~L.,  {Rowe} J.~F.,
  {Ravichandran} G.,  {Petigura} E.~A.,  {Haas} M.~R.,   {Batalha} N.~M.,
  2016, \mn@doi [ApJ] {10.3847/0004-637X/822/2/86}, \href
  {http://adsabs.harvard.edu/abs/2016ApJ...822...86M} {822, 86}

\bibitem[\protect\citeauthoryear{{Murray} \& {Dermott}}{{Murray} \&
  {Dermott}}{1999}]{Murray-1999}
{Murray} C.~D.,  {Dermott} S.~F.,  1999, Solar system dynamics

\bibitem[\protect\citeauthoryear{{Oza} et~al.,}{{Oza}
  et~al.}{2019}]{Oza-etal-2019}
{Oza} A.~V.,  et~al., 2019, \mn@doi [ApJ] {10.3847/1538-4357/ab40cc}, \href
  {https://ui.adsabs.harvard.edu/abs/2019ApJ...885..168O} {885, 168}

\bibitem[\protect\citeauthoryear{{Quarles}, {Li}  \&
  {Rosario-Franco}}{{Quarles} et~al.}{2020}]{Quarles-Rosario-Franco-2020}
{Quarles} B.,  {Li} G.,   {Rosario-Franco} M.,  2020, \mn@doi [ApJl]
  {10.3847/2041-8213/abba36}, \href
  {https://ui.adsabs.harvard.edu/abs/2020ApJ...902L..20Q} {902, L20}

\bibitem[\protect\citeauthoryear{{Rein} \& {Spiegel}}{{Rein} \&
  {Spiegel}}{2015}]{Rein-Spiegel-2015}
{Rein} H.,  {Spiegel} D.~S.,  2015, \mn@doi [MNRAS] {10.1093/mnras/stu2164},
  \href {http://adsabs.harvard.edu/abs/2015MNRAS.446.1424R} {446, 1424}

\bibitem[\protect\citeauthoryear{{Rosario-Franco}, {Quarles}, {Musielak}  \&
  {Cuntz}}{{Rosario-Franco} et~al.}{2020}]{Rosario-Franco-etal-2020}
{Rosario-Franco} M.,  {Quarles} B.,  {Musielak} Z.~E.,   {Cuntz} M.,  2020,
  \mn@doi [AJ] {10.3847/1538-3881/ab89a7}, \href
  {https://ui.adsabs.harvard.edu/abs/2020AJ....159..260R} {159, 260}

\bibitem[\protect\citeauthoryear{{S{\'a}nchez}, {de El{\'\i}a}  \&
  {Downes}}{{S{\'a}nchez} et~al.}{2020}]{Sanchez-etal-2020}
{S{\'a}nchez} M.~B.,  {de El{\'\i}a} G.~C.,   {Downes} J.~J.,  2020, \mn@doi [A
  \& A] {10.1051/0004-6361/201937317}, \href
  {https://ui.adsabs.harvard.edu/abs/2020A&A...637A..78S} {637, A78}

\bibitem[\protect\citeauthoryear{{Teachey}}{{Teachey}}{2021}]{Teachey-2021}
{Teachey} A.,  2021, \mn@doi [MNRAS] {10.1093/mnras/stab1840}, \href
  {https://ui.adsabs.harvard.edu/abs/2021MNRAS.506.2104T} {506, 2104}

\bibitem[\protect\citeauthoryear{Teachey \& Kipping}{Teachey \&
  Kipping}{2018}]{Teachey-Kipping-2018}
Teachey A.,  Kipping D.~M.,  2018, \mn@doi [Science Advances]
  {10.1126/sciadv.aav1784}, 4

\bibitem[\protect\citeauthoryear{{Teachey}, {Kipping}  \& {Schmitt}}{{Teachey}
  et~al.}{2018}]{Teachey-etal-2018}
{Teachey} A.,  {Kipping} D.~M.,   {Schmitt} A.~R.,  2018, \mn@doi [AJ]
  {10.3847/1538-3881/aa93f2}, \href
  {http://adsabs.harvard.edu/abs/2018AJ....155...36T} {155, 36}

\bibitem[\protect\citeauthoryear{{Teachey}, {Kipping}, {Burke}, {Angus}  \&
  {Howard}}{{Teachey} et~al.}{2020}]{Teachey-etal-2020}
{Teachey} A.,  {Kipping} D.,  {Burke} C.~J.,  {Angus} R.,   {Howard} A.~W.,
  2020, \mn@doi [AJ] {10.3847/1538-3881/ab7001}, \href
  {https://ui.adsabs.harvard.edu/abs/2020AJ....159..142T} {159, 142}

\bibitem[\protect\citeauthoryear{{Tokadjian} \& {Piro}}{{Tokadjian} \&
  {Piro}}{2020}]{Tokadjian-Piro-2020}
{Tokadjian} A.,  {Piro} A.~L.,  2020, \mn@doi [AJ] {10.3847/1538-3881/abb29e},
  \href {https://ui.adsabs.harvard.edu/abs/2020AJ....160..194T} {160, 194}

\bibitem[\protect\citeauthoryear{{Wisdom}}{{Wisdom}}{1980}]{Wisdom-1980}
{Wisdom} J.,  1980, \mn@doi [AJ] {10.1086/112778}, \href
  {https://ui.adsabs.harvard.edu/abs/1980AJ.....85.1122W} {85, 1122}

\bibitem[\protect\citeauthoryear{{Zollinger}, {Armstrong}  \&
  {Heller}}{{Zollinger} et~al.}{2017}]{Zollinger-etal-2017}
{Zollinger} R.~R.,  {Armstrong} J.~C.,   {Heller} R.,  2017, \mn@doi [MNRAS]
  {10.1093/mnras/stx1861}, \href
  {http://adsabs.harvard.edu/abs/2017MNRAS.472....8Z} {472, 8}

\makeatother
\end{thebibliography}
